\documentclass[a4paper,12pt]{article} 

\usepackage [T1]{fontenc}
\usepackage {calligra}
\usepackage [T1]{pbsi}
\usepackage{CJKutf8} 
\usepackage[colorlinks=true,linkcolor=blue]{hyperref}
\usepackage{amsmath,amssymb,amsfonts} 
\usepackage{graphicx}                 
\usepackage{color}                    
\usepackage{hyperref}                 
\usepackage{cite}
\definecolor{red}{rgb}{1,0,0}           
\definecolor{green}{rgb}{0,1,0}
\definecolor{blue}{rgb}{0,0,1}
\definecolor{darkblue}{rgb}{0,0,0.5}
\definecolor{lightblue}{rgb}{.5,.5,1}
\definecolor{lightgray}{gray}{.87}          
\definecolor{Dark}{gray}{.20}
\definecolor{pink}{rgb}{.95,0.82,0.92}  
\definecolor{yellow}{rgb}{1,1,0}
\definecolor{lightyellow}{rgb}{1,1,.5}
\definecolor{purple}{rgb}{0.7,0,0.85}
\definecolor{darkgreen}{rgb}{0,0.45,0}
\definecolor{orange}{rgb}{0.8,0.2,0.2}
\def \be {\begin{equation}}
\def \ee {\end{equation}}
\def \bea {\begin{eqnarray}}
\def \eea {\end{eqnarray}}
\def \nn {\nonumber}

\def \rr {\raise.35ex\hbox{\small $\prime$}\kern-.17em{\mbox{\large $\imath$}}}
\def \del {\partial}
\def \dels {\partial\kern-.5em / \kern.5em}
\def \As {{A\kern-.5em / \kern.5em}}
\def \Ds {D\kern-.7em / \kern.5em}

\def \a {\alpha}

\def \b {\beta}

\def \g {\gamma}
\def \G {\Gamma}
\def \d {\delta}
\def \eps {\epsilon}

\def \lam {\lambda}
\def \Lam {\Lambda}
\def \s {\sigma}

\def \om {\omega}

\def \th {\theta}

\def \Gb {{\bf\Gamma}}
\def \Tb {{\bf T}}
\def \Rb {{\bf R}}

\def \Sb {{\bf S}}
\def \Kb {{\bf K}}
\def \Ub {{\bf U}}

\newcommand{\solution}[1]{}

\newcommand{\detail}[1]{}

\newcommand{\hide}[1]{}

\begin{document}

\pagestyle{plain}

\begin{CJK}{UTF8}{bsmi} 

\begin{titlepage}


\begin{center}

\noindent
\textbf{\LARGE
Geometry of Area Without Length 
}

\vskip .5in
{\large 
Pei-Ming Ho$\,{}^{a,b}$\footnote{e-mail address: pmho@phys.ntu.edu.tw},
Takeo Inami$\,{}^{a,c}$\footnote{email address: inami@phys.chuo-u.ac.jp}
}
\\

{\vskip 10mm \sl
${}^a$
Department of Physics and Center for Theoretical Sciences, \\
National Taiwan University, Taipei 106, Taiwan, R.O.C. \\
${}^b$
Center for Advanced Study in Theoretical Sciences, \\
National Taiwan University, Taipei 106, Taiwan, R.O.C. \\
${}^c$
Riken Bishina Center, Saitam, Japan
}\\
\vskip 3mm
\vspace{60pt}

\begin{abstract}

To define a free string by the Nambu-Goto action,
all we need is the notion of area,
and mathematically
the area can be defined directly in the absence of a metric.
Motivated by the possibility that 
string theory admits backgrounds 
where the notion of length is not well defined
but a definition of area is given,
we study space-time geometries based on 
the generalization of metric to {\em area metric}.
In analogy with Riemannian geometry,
we define the analogues of connections,
curvatures and Einstein tensor.
We propose a formulation generalizing
Einstein's theory that will be useful 
if at a certain stage or a certain scale 
the metric is ill-defined and 
the space-time is better characterized by the notion of area.
Static spherical solutions are found for the generalized Einstein equation in vacuum,
including the Schwarzschild solution as a special case.

\end{abstract}

\end{center}

\end{titlepage}

\setcounter{page}{1}
\setcounter{footnote}{0}
\setcounter{section}{0}


\section{Introduction}

The string perturbation theory is fully specified
by the 2-dimensional world-sheet theory of a free string,
which is classically defined by the Nambu-Goto action 
in a purely geometric background.
As the Nambu-Goto action simply equals the world-sheet area
(up to an overall factor of the string tension),
it is natural to speculate the possibility that 
string theory admits space-time manifolds
on which only the notion of area is defined, 
in the absence of a metric to define the notion of length.
This topic is also interesting from a purely mathematical point of view.

It may be surprising that,
in fact, 
not only the Nambu-Goto action,
but also the Yang-Mills action
\be
S = \frac{1}{4} \int d^D x\; \mbox{Tr}(F_{ij} g^{ik}g^{jl} F_{kl})
= \frac{1}{8} \int d^D x\; \mbox{Tr}(F_{ij} h^{ijkl} F_{kl}),
\ee
depends only on the combination
$h_{ijkl} = g_{ik}g_{jl}-g_{il}g_{jk}$
(and its inverse)
that defines area (see below),
and is not directly sensitive to the metric $g_{ij}$. 
In particular,
Maxwell's theory on a manifold with a generic area metric $h_{ijkl}$
which cannot be associated with any metric $g_{ij}$
is motivated by the study of electromagnetic phenomenon 
in a generic linear media.
Birefringence effect in crystal,
and quantum loop effect in curved space-time \cite{Drummond:1979pp}
can both lead to a more general area metric $h_{ijkl}$
than those defined by metrics.

Perhaps the notion of area is more fundamental 
than that of length in physics
from the perspectives of string theory.
Basic notions of the geometry for area metric $h_{ijkl}$
(assuming that it is not defined by any metric $g_{ij}$),
including area connection, area torsion and area curvature,
have been studied in the literature \cite{Maran:2005iu}--\cite{Dahl:2011hb}.
The aim of this paper is to write down an equation for the area metric 
analogous to the Einstein equation.
In fact, 
an analogue of the Einstein equation was given in Ref.\cite{Punzi:2006nx},
with the help of an effective metric
defined from the area metric.
Instead we hope to deal with geometric structures 
which cannot be properly characterized by any metric (or effective metric),
and we would like to provide an alternative approach 
in which the area metric will be the only quantity that 
defines geometry in our formulation.
This means that we need to find a way to 
associate a particular area connection to any given area metric.
Such an area connection was not discussed in the past.

The plan of the paper is the following.
In Sec. \ref{notion-area} 
we point out two scenarios in which 
area metric geometry potentially arises 
as an alternative theory of gravity.
The focus of the paper is the scenario of string theory,
on which we have already briefly commented above.
In Sec. \ref{world-sheet}
we introduce the notion of area metric,
and explore its algebraic properties.
We note in Sec. \ref{string-action}
that the Nambu-Goto action for an area metric 
is equivalent to a superposition of Polyakov actions 
for a set of metrics.

The notion of area connection is introduced in Sec. \ref{area-connection}.
Contrary to the uniqueness of the Levi-Civita connection in Riemannian geometry,
the area metricity condition and the area torsion-free condition 
are not sufficient to uniquely fix the area connection,
because there are a lot more covariant degrees of freedom 
in the area connection.
We show in Sec. \ref{GGE} that,
assuming the area metricity condition and the area torsion-free condition,
the equation of motion derived from 
the Nambu-Goto action for a generic area metric
can be interpreted as 
the condition for the tangent area of the world-sheet
to be parallel transported along the tangent space.

In Sec. \ref{more-connection},
we design a procedure to modify any area connection 
by adding a tensor such that it becomes area torsion-free,
and an analogue of the Levi-Civita connection is defined
not only to satisfy the area metricity condition 
and the area torsion-free condition,
but also to minimize its covariant degrees of freedom.
Area curvature is defined in Sec. \ref{area-curvature},
and we find a generalization of Einstein's equation in vacuum
for the area metric.
In Sec. \ref{example},
we focus on the examples of area metrics which are diagonalizable.
We find static, spherical solutions to 
the generalized Einstein equation in vacuum in Sec. \ref{spherical},
including the Schwarzschild solution and other solutions 
that can or cannot be associated to ordinary metrics.
Finally, 
we summarise and comment in the last section.

\section{Area vs length}
\label{notion-area}

In physics we almost always assume that the space-time 
is a pseudo-Riemannian manifold equipped with the notion of length
\be
ds^2 = g_{ij}(x) dx^{i} \otimes dx^{j},
\label{ds=g}
\ee
where $g_{ij}$ is a symmetric tensor called metric.
This determines the action for a free particle under the effect 
of nothing but the space-time geometry (gravity) 
to be given by the length of the particle's world-line
\be
S = m \int ds = m \int d\tau \; \sqrt{g_{ij} \dot{x}^{i} \dot{x}^{j}},
\label{S=L}
\ee
where $\dot{x}^{i} \equiv \frac{d}{d\tau}x^{i}$ and
$i, j = 0, 1, 2, \cdots, (D-1)$ in a $D$-dimensional space-time.
The coefficient $m$ is identified with the mass of the particle.
By observing the behavior of a free particle,
one can infer properties of the space-time.
Furthermore,
since the the action (\ref{S=L}) also determines how 
interactions mediated by particles propagate in space-time,
the notion of length defined by the free particle action (\ref{S=L})
is expected to play an important role in all physics of particles.
For example, 
the magnitude of the force for an interaction
should depend on the distance
between two particles.

Mathematically, 
in addition to the Riemannian structure, 
there are other geometrical notions 
that can be defined on a manifold.
In particular, 
one can consider a manifold equipped with 
the notion of area but not the notion of length.
Analogous to (\ref{ds=g}),
we will assume that the infinitesimal area element $da$
can be defined through the expression
\be
da^2 \equiv 
\frac{1}{4} h_{ijkl}(x) (dx^i \wedge dx^j) \, \otimes \, (dx^k \wedge dx^l),
\ee
where the tensor $h_{ijkl}(x)$ will be referred to 
as the {\em area metric}.
If a Riemannian metric $g_{ij}$ is defined,
the area metric should be given by
\bea
h_{ijkl} 
&\equiv&
g_{ik} g_{jl} - g_{il} g_{jk}.
\label{h=gg}
\eea
Yet the area metric can be defined 
in the absence of a metric,
for a manifold on which the notion of length is not defined.
It is possible that
the universe is at a certain stage or a certain scale 
characterized by the notion of area instead of the notion of length.

There are at least two scenarios in which 
the geometry of space-time is characterized 
not by the notion of length, 
but by the notion of area. 
The first scenario occurs in the context of 
quantum gravity
when the quantum state of the space-time is
a superposition of several coherent states 
with different metrics.
The second scenario happens in string theory
as a more general background 
for the Nambu-Goto action of a free string.

First, 
in a quantum theory of gravity,
a classical space-time metric $g_{ij}(x)$ 
can be interpreted as the expectation value of 
the metric operators $\hat{g}_{ij}(x)$ 
for a coherent state $|g\rangle$.
It is possible that the expectation value of $\hat{g}_{ij}(x)$ vanishes
\be
g_{ij}(x) \equiv \langle \Psi | \hat{g}_{ij}(x) | \Psi \rangle = 0
\label{g=0}
\ee
for a state $|\Psi\rangle$,
while the expectation value of the area metric 
\be
\hat{h}_{ijkl} \equiv
\hat{g}_{ik}(x)\hat{g}_{jl}(x)
- \hat{g}_{il}(x)\hat{g}_{jk}(x)
\ee
is non-zero:
\be
h_{ijkl}(x) \equiv
\langle \Psi | \hat{h}_{ijkl} | \Psi \rangle 
\neq 0.
\label{h-hhat}
\ee
In this space-time background $|\Psi\rangle$,
the notion of length is trivial
(vanishing distance between any two points),
and the notion of area is expected to play 
a more important role in the formulation of physical laws.

To be more concrete,
let $| g^{(\a)} \rangle$ denote the eigenstates of 
the operator $\hat{g}_{ij}(x)$
with the eigenvalues $g_{ij}^{(\a)}(x)$.
As eigenstates of Hermitian operators, 
the states $| g^{(\a)} \rangle$ are orthogonal to each other.
For a superposition of eigenstates
\be
| \Psi \rangle \equiv \sum_{\a} | g^{(\a)} \rangle,
\label{Psi=sum}
\ee
we have
\bea
g_{ij} &=& \sum_{\a} g_{ij}^{(\a)},
\\
h_{ijkl}(x) &=& 
\sum_{\a} \left[g^{(\a)}_{ik}(x) g^{(\a)}_{jl}(x) - g^{(\a)}_{il}(x) g^{(\a)}_{jk}(x) \right].
\label{hath}
\eea
For instance,
if $g^{(1)} = - g^{(2)}$ $(\a = 1, 2)$,
we have vanishing expectation value of the metric $\hat{g}$ (\ref{g=0})
and the area metric (\ref{h-hhat}) is determined 
by a metric $g = \sqrt{2} g^{(1)}$ through (\ref{h=gg}).
In this case, 
even though the area metric is given by a specific metric $g$, 
we do not expect the metric $g$ to couple to matters in the usual way,
because it is not the expectation value of the metric operator.

In general,
there are infinitely many sets of metrics $\{g^{(\a)}\}$
that can be related to the same $h$ through (\ref{hath}).
A simple way to see this is the following.
Assuming that (\ref{hath}) holds for a set of metrics $\{g^{(\a)}\}_{\a=1}^{n}$,
we find that $\{g'{}^{(\a)}\}_{\a=1}^{n}$ also satisfies (\ref{hath})
if $g'{}^{(\a)}_{ij} = M^{\a}{}_{\b}g^{(\b)}_{ij}$
for any $SO(n)$ matrix $M^{\a}{}_{\b}$ of functions.

In this scenario, 
the quantum state $\Psi$ (\ref{Psi=sum}) is different
for different choices of the set of metrics $\{g^{(\a)}\}$.
Hence two sets of metrics $\{g^{(\a)}\}$ and $\{g'{}^{(\a)}\}$ are physically inequivalent even if 
the area metric $h_{ijkl}$ (\ref{hath}) is exactly the same.
The geometric information is not fully characterized by the area metric,
but is by the set of metrics $\{g^{(\a)}\}$.

The second scenario in which area replaces length
as the fundamental geometric notion of space-time
occurs in string theory.
In string theory,
forces are mediated by strings,
and the only necessary notion of geometry 
for the string world-sheet action
is the area,
not the length.
While a metric always defines an area metric,
{\em a priori}
a well-defined area does not necessarily imply a unique well-defined metric.
Hence we expect that, 
in string theory,
it is sensible to consider space-time background geometries 
characterized by the notion of area,
in the absence of the notion of length.
In the background in which the string world-sheet action 
is fully determined by an area metric,
the complete information of space-time geometry 
is encoded in the area metric.
The second scenario will be the focus of this work,
although most of the discussions below
apply to both scenarios.

The main difference between these two scenarios 
is that in the first scenario the geometry is 
characterized by a set of metrics $\{g^{(\a)}\}$,
while in the second scenario the full geometry is 
encoded in the area metric alone.
In this paper we focus on the second scenario (string theory),
and we will comment in Sec.\ref{effective-metric} on viable mechanisms
to explain why the current universe 
can be described as a Riemannian manifold
to a good approximation.

\section{Area metric}
\label{world-sheet}

In this section we introduce basic notions about the area metric.

The area of a surface $\Sigma$
is defined as the integral of
the infinitesimal area element:
\be
\mbox{Area} = \int_\Sigma da,
\label{int-da-1}
\ee
The surface can be defined by 
its embedding coordinates $X^i(\s)$
in the $D$-dimensional space,
with $\{\s^0$, $\s^1\}$ an arbitrary coordinate system on $\Sigma$.
When the space-time is equipped with a metric $g_{ij}$,
one can define the induced metric $G_{ab}$
on the surface $\Sigma$ as
\be
G_{ab}
\equiv g_{ij}(X) \del_{a}X^{i}(\s) \del_{b}X^{j}(\s),
\label{induced-G}
\ee
where $a, b = 0, 1$ label the surface directions,
and an infinitesimal area element of the world-sheet is given by
\be
da = d^2\s \; \sqrt{\det G}.
\ee
More explicitly, 
it is
\bea
da^2 
=
\frac{1}{4} \eps^{ab}\del_{a}X^{i}\del_{b}X^{j} 
h_{ijkl}
\eps^{cd}  \del_{c}X^{k}\del_{d}X^{l} |d^2 \s|^2
=
\frac{1}{4} A^{ij} h_{ijkl} A^{kl},
\label{AhA}
\eea
where $h_{ijkl}$ is given by (\ref{h=gg}) and
\bea
A^{ij} 
&\equiv&
\eps^{ab} \; \del_{a}X^{i}\del_{b}X^{j}\, (d\s^0 \wedge d\s^1) 
= dX^i \wedge dX^j.
\label{A=dXdX}
\eea
The last expression (\ref{AhA}) is in the form of 
a norm on the space of two-forms $dX^i\wedge dX^j$.

A natural and modest way to generalize the notion of area is to relax $h_{ijkl}$ 
from the definition (\ref{h=gg}) based on a given metric $g_{ij}$.
It is possible to generalize the notion of area further, 
just like the notion of length can be generalized in Finsler geometry.
Yet, 
as a first step, 
we restrict ourselves to the notion of area defined by 
a norm in the quadratic form (\ref{AhA}),
where $A^{ij}$ is given by (\ref{A=dXdX}) but 
$h_{ijkl}$ does not have to be given by (\ref{h=gg}).
(But it will be restricted by other constraints.)
We will see below that 
this definition of area is suitable for the application to string theory.

Notice that not all components of a rank-4 tensor $h_{ijkl}$ 
are relevant in the definition of the area metric
\be
da^2 \equiv \frac{1}{4} h_{ijkl} A^{ij} A^{kl}.
\label{da=h}
\ee
We should impose conditions to remove 
those irrelevant components of $h_{ijkl}$.
First, it is obvious that
we should impose the symmetry properties
\be
h_{ijkl} = - h_{jikl}
= - h_{ijlk} = h_{klij}.
\label{h-symm-1}
\ee
It is less obvious that
the area metric should also satisfy the {\em cyclicity condition}
\cite{Schuller:2005yt}
\be
h_{ijkl} + h_{jkil} + h_{kijl} = 0.
\label{h-symm-2}
\ee
Due to the cyclicity condition (\ref{h-symm-2}),
the definition of area metric is different from that of a generic norm for 2-forms.
The necessity of the cyclicity condition can be seen by rewriting (\ref{AhA}) as
\be
da^2 = \hat{h}_{(ij)(kl)} \dot{X}^i \dot{X}^j X^{\prime k} X^{\prime l} \; |d\s^0 d\s^1|^2,
\ee
where derivatives of $\s^0$ are denoted by dots, 
and derivatives of $\s^1$ by primes,
and
\be
\hat{h}_{(ij)(kl)} \equiv 
h_{ikjl}+h_{jkil}
\label{h-hat}.
\ee
In other words,
only the tensor $\hat{h}_{ijkl}$ (\ref{h-hat}) is necessary to define the area,
and the cyclicity condition removes the irrelevant part in $h_{ijkl}$.

\subsection{Algebraic properties of area metric}
\label{algebra-metric}

\subsubsection{Gilkey decomposition}

Let us count the number of free parameters in $h$
subject to the constraints (\ref{h-symm-1}), (\ref{h-symm-2}).
In space-time of $D = d+1$ dimensions,
the number of independent components in $h_{ijkl}$ is
\be
\frac{C^D_2(C^D_2+1)}{2}-C^D_4 = \frac{1}{12}D^2(D-1)(D+1).
\label{NumDoF-h}
\ee

In 2D, 
$g_{ij}$ has 3 independent components, 
while $h_{ijkl}$ has only one.
The area metric can always be put in the form (\ref{h=gg})
for infinitely many choices of $g$,
as there are more independent free components in $g$ than $h$.
It is less interesting to talk about the area metric in 2D
because all area metrics are locally equivalent to $h_{1212} = \pm 1$.

In 3D, 
both $g_{ij}$ and $h_{ijkl}$ have 6 independent components.
The metric $g_{ij}$ can always be determined for a given $h_{ijkl}$ up to a $\pm$ sign
through the  relation (\ref{h=gg}).
The difference between metric and area metric is minimal.

In higher dimensions,
$h_{ijkl}$ is in general not of the form (\ref{h=gg}).
There are a lot more free parameters in $h$ than $g$.
In general,
according to a theorem by Gilkey \cite{Gilkey},
the properties (\ref{h-symm-1}) and (\ref{h-symm-2}) imply that
the area metric can always be expressed in the form
\be
h_{ijkl} \equiv
\sum_{\a}
\left[
g^{(\a)}_{ik} g^{(\a)}_{jl} - g^{(\a)}_{il} g^{(\a)}_{jk}
\right]
\label{h=gg-A}
\ee
for a set of metrics $g^{(\a)}_{ij}$.
Thus we can always think of the area metric as 
the superposition of a set of metrics.
This expression (\ref{h=gg-A}) fits very well
with the physical interpretation
of the area metric as the effect of
a superposition of states of different metrics (\ref{hath}).
On the other hand,
the decomposition of a given area metric $h$ 
into the superposition of area metrics for different metrics
is not unique.
There are always infinitely many sets of $\{g^{(\a)}\}$
for the same area metric $h$.

\subsubsection{The inverse $h^{-1}$}

We define another tensor $h^{ijkl}$ as the inverse of $h_{ijkl}$ by
\be
\frac{1}{2} h^{ijkl}h_{klmn} =
\d^{i}_{m}\d^{j}_{n} - \d^{i}_{n}\d^{j}_{m}.
\ee
We use the convention that when we sum over 
a pair of (anti-) symmetrized indices $[kl]$
we insert a factor of $1/2$
because the labels $[ij]$ and $[ji]$ are redundant.
The tensor on the right hand side of the equation,
\be
I^{ij}_{kl} \equiv
\d^i_k\d^j_l-\d^j_k\d^i_l,
\ee
plays the role of identity as a matrix with
indices that are pairs of anti-symmetrized indices $[ij]$.
It satisfies
\be
\frac{1}{2} I^{ij}_{kl} I^{kl}_{mn} = I^{ij}_{mn}.
\ee

For $h$ given by (\ref{h=gg}),
$h^{ijkl} = g^{ik}g^{jl}-g^{il}g^{jk}.$

\subsubsection{Volume form}

The area metric $h_{ijkl}$ can be viewed 
as a symmetric $N\times N$ matrix for $N = C^D_2$
in $D$ dimensions.
The determinant $(\det h)$ of this matrix 
is of dimension $[L]^{4N}$ since $h$ is of dimension $[L]^4$,
where $[L]$ represents the dimension of length.
The determinant $(\det h)$ can be defined by
\be
(\det h) \eps_{A_1 \cdots A_N} =
\eps^{B_1 \cdots B_N}h_{A_1 B_1}\cdots h_{A_N B_N},
\label{deth}
\ee
where the indices $A_i, B_i$ represent 
pairs of anti-symmetrized indices $[ij]$ ($i, j = 0, 1, \cdots, D-1$),
with $[ij]$ and $[ji]$ identified,
and $\eps^{B_1\cdots B_N}$
is the totally anti-symmetrized tensor of rank $N$.

In our convention,
a sum over a pair of anti-symmetrized indices $A=[ij]$
is understood as a sum over a pair of ordered indices.
For example,
in 4 dimensions
with $i,j=0,1,2,3$,
a pair of indices $[ij]$ can take 12 different values
with $i\neq j$.
But the indices $A, B$ take only 6 values
$[01],[02],[03],[12],[13],[23]$ in the summation.
In general,
we have
\be
\Phi^A \Psi_A = \frac{1}{2} \Phi^{[ij]} \Psi_{[ij]}.
\ee

In order to develop differential calculus in space-time,
it is desirable to define the volume form
\be
\Omega_{i_1\cdots i_D} \equiv 
\om \eps_{i_1\cdots i_D},
\label{Omega}
\ee
where
\be
\om \equiv (\det h)^{1/(2(D-1))}
\ee
is of dimension $[L]^{D}$.
This reminds us that
there are two totally anti-symmetrized tensors
\be
\eps_{i_1\cdots i_D}
\quad \mbox{and} \quad
\eps_{A_1\cdots A_N},
\ee
where $i_1, \cdots, i_D = 0, 1, \cdots, D-1$
and $A_1, \cdots, A_N = 1, 2, \cdots, N$.
We have defined the determinant of the area metric
using $\eps_{A_1\cdots A_N}$,
but we can also define another determinant, 
denoted $\det{}'h$, 
for even $D$, as
\be
(\det{}' h) 
= \frac{1}{(2^n)^2 n!} \, \eps^{[i_1 j_1]\cdots[i_n j_n]} \eps^{[k_1 l_1]\cdots[k_n l_n]} \,
h_{[i_1 j_1][k_1 l_1]} \cdots h_{[i_n j_n][k_n l_n]}
\ee
where $n \equiv D/2$.

A natural question is whether the two determinants
are related to each other in a simple way.
In the special case when
the area metric is defined by a metric (\ref{h=gg}),
we have $\det' h = \det g$ and
\be
\det h = (\det{}' h)^{D-1}.
\label{special-h}
\ee
In general,
the two determinants are algebraically independent.

\subsubsection{Effective metric}
\label{effective-metric}

Given an area metric $h$,
we would like to know when it can be defined
by a metric $g$ through (\ref{h=gg}).
We hope to find conditions on $h$ that tell us 
whether a metric $g$ exists for (\ref{h=gg}) to hold.
We discuss this issue
separately for even and odd dimensions.

\noindent{\bf Odd dimensions}

For $D = 2n +1$ ($n > 0$),
let 
\be
H^{ij} \equiv \frac{1}{2^n (2n)!} (\det h)^{-1/(2n)} 
\eps^{ii_1\cdots i_{2n}}\eps^{jj_1\cdots j_{2n}}
h_{i_1 i_2 j_1 j_2} \cdots h_{i_{2n-1} i_{2n} j_{2n-1} j_{2n}},
\ee
so that if $h$ is defined from a metric $g$ (\ref{h=gg}),
one can choose to define $g$ by
\be
g^{ij} = \pm H^{ij}.
\ee

For odd dimensions, 
it is tempting to define $H$
as the effective metric for a given $h$,
and to use it to define Levi-Civita connection and Riemannian curvature,
even when (\ref{h-test}) is not satisfied.
However, we do not expect it to be sufficient to
fully characterize the geometry
because there are similar inequivalent expressions
which are equally justified to be called the effective metric.
For example,
we could start with an inverse of $h$ 
and combine it with the $\eps$ tensor as
\be
H'_{ij} \equiv \frac{1}{2^n (2n)!} (\det h)^{1/(2n)} 
\eps_{ii_1\cdots i_{2n}}\eps_{jj_1\cdots j_{2n}}
h^{i_1 i_2 j_1 j_2} \cdots h^{i_{2n-1} i_{2n} j_{2n-1} j_{2n}},
\ee
which is in general not the inverse of $H^{ij}$.

Furthermore,
the effective metric is in general not covariantly constant
because the volume form is not,
regardless of how one defines the area connection,
as we will see below in Appendix \ref{app}.

Define
\bea
\Delta^{ijkl} &\equiv& h^{ijkl} - (H^{ik}H^{jl} - H^{il}H^{jk}),
\\
\Delta'_{ijkl} &\equiv& h_{ijkl} - (H^{-1}_{ik}H^{-1}_{jl} - H^{-1}_{il}H^{-1}_{jk}).
\eea
Both of them vanish if (\ref{h=gg}) holds for some metric $g$.
Conversely, both
\be
\Delta^{ijkl}=0
\quad \mbox{and} \quad
\Delta'_{ijkl}=0
\label{h-test}
\ee
are sufficient to ensure that
the area metric can be derived from a metric.
Introducing a potential energy $V(\Delta, \Delta')$
that is minimized at $\Delta=0$ or $\Delta'=0$ 
provides a mechanism that 
drives the background geometry towards
one that can be fully defined by a metric.

\noindent{\bf Even Dimensions}

For $D=2n+2$,
let
\be
H^{ijkl} \equiv \frac{1}{2^n (2n)!} (\det h)^{-1/(2n+1)} 
\eps^{iji_1\cdots i_{2n}}\eps^{klj_1\cdots j_{2n}}
h_{i_1 i_2 j_1 j_2} \cdots h_{i_{2n-1} i_{2n} j_{2n-1} j_{2n}}.
\ee
This tensor $H^{ijkl}$ agrees with the inverse of $h_{ijkl}$
when $h_{ijkl}$ is defined by a metric $g_{ij}$ (\ref{h=gg}).
Similarly one can define
\be
H'_{ijkl} \equiv \frac{1}{2^n (2n)!} (\det h)^{1/(2n+1)} 
\eps_{iji_1\cdots i_{2n}}\eps_{klj_1\cdots j_{2n}}
h^{i_1 i_2 j_1 j_2} \cdots h^{i_{2n-1} i_{2n} j_{2n-1} j_{2n}}.
\ee
If $h_{ijkl}$ is defined by a metric $g$ (\ref{h=gg}),
we have both conditions
\bea
H^{ijkl}-h^{ijkl}=0
\quad \mbox{and} \quad
H'_{ijkl}-h_{ijkl}=0
\eea
satisfied.
In 4D,
it can be checked explicitly that
these conditions are strong enough to imply the existence 
of a metric for eq.(\ref{h=gg}) to hold
only for some of the meta-classes defined in Ref.\cite{Schuller:2009hn}.
We expect similar situation for higher dimensions.

One can define
\bea
\Delta^{ijkl} \equiv H^{ijkl}-h^{ijkl},
\qquad
\Delta'_{ijkl} \equiv H'_{ijkl}-h_{ijkl},
\eea
and a potential energy $V(\Delta, \Delta')$
minimized at $\Delta = \Delta' = 0$ would provide 
a mechanism to enhance the possibility of 
finding our universe evolving towards a state 
approximately described by a metric.
But even when $\Delta = \Delta' = 0$,
it is possible that (\ref{h=gg}) does not hold for any metric.
In the geometry of area metric,
even dimensions and odd dimensions are 
quite different in this aspect.

\subsection{String world-sheet action}
\label{string-action}

The Nambu-Goto action is defined by the area of the world-sheet
\be
S = T_s \int da = T_s \int d^2 \s \; {\cal A}_h,
\label{generalized-NG}
\ee
where $T_s$ is the tension of the string,
and
\be
{\cal A}_h^2 \equiv 
\frac{1}{4} \eps^{ab}\eps^{cd}
h_{ijkl}(X)
\del_{a}X^{i}(\s)\del_{b}X^{j}(\s)\del_{c}X^{k}(\s)\del_{d}X^{l}(\s).
\label{H}
\ee
This action is invariant under both world-sheet diffeomorphism 
and space-time diffeomorphism.
It is well defined regardless of whether
$h_{ijkl}$ is defined from a metric through (\ref{h=gg}) or not.
From the viewpoint of strings,
the notion of length is unnecessary.
It is more natural to start with a space-time geometry
defined by the area metric.

In the perturbative string theory,
the dynamics of the background fields
can be derived from the string world-sheet theory
by requiring conformal symmetry,
but one has to first rewrite the Nambu-Goto action in another way
before the quantization of the theory can be carried out.

The Nambu-Goto action can be written in various different ways.
It can be shown to be equivalent to the action
\be
S' = \frac{T_s}{2} \int d^2 \s \; [e^{-1}(\s) {\cal A}_h^2 + e(\s)],
\ee
where $e(\s)$ represents the measure of area on the world sheet.
Solving the equation of motion for $e(\s)$ 
reproduces the action $S$.
Despite the fact that the square root
is removed in this action,
it is still hard to quantize due to its quartic form.

Another way to rewrite the action is the following.
As the area metrics can always be defined 
by a superposition of metrics $g^{(\a)}$ (\ref{h=gg-A}),
the Nambu-Goto action (\ref{generalized-NG}) 
can always be put in the form
\be
S = T_s \int d^2 \s \; \sqrt{\sum_{\a=1}^{n} \det G^{(\a)}},
\ee
where $G_{\a\b}^{(\a)}$ is the induced metric
\be
G^{(\a)}_{ab} \equiv g^{(\a)}_{ij} \del_{a}X^i \del_{b}X^j.
\ee
It is equivalent to the following action quadratic in $X$:
\be
S'' = \frac{T_s}{2} \int d^2 \s \;  \sqrt{\g} 
\left[
\sum_{\a=1}^{n} (\g^{(\a)})^{-1 ab} g^{(\a)}_{ij} \del_{a}X^i \del_{b}X^j
\right],
\ee
where 
\be
\g^{-1} \equiv \sum_{\b=1}^{n}(\det \g^{(\b)})^{-1}
\ee
and $n$ is the number of metrics $g^{(\a)}_{ij}$ appearing in (\ref{h=gg-A}).
This action is a superposition of 
the world-sheet actions for the space-time metrics $g^{(\a)}$,
with a superposition of the world-sheet metrics $\g^{(\a)}$ accordingly.
This action has both world-sheet and space-time diffeomorphism symmetries, 
as well as the Weyl symmetry: 
\be
\g^{(\a)}_{ab} \rightarrow \g^{(\a)\prime}_{ab} = e^{\phi(\s)} \g^{(\a)}_{ab}.
\ee

We will leave the quantization of this action
for future works.
Instead we make some comments here.
First we recall that
the Nambu-Goto action of a string 
in a generic Riemannian space-time 
is classically equivalent to the Polyakov action, 
which can be quantized as a perturbation theory
around the Minkowski background.
Einstein's equation is the constraint 
on the background at the leading order
to ensure that the beta functions vanish.
In principle, 
this procedure of deriving
Einstein's equation from the string world-sheet action
can be generalized to derive a field equation 
for the area metric.
We hope that,
even without an actual calculation of the beta functions,
it is possible to have a good guess about
the generalized Einstein's equation
if we know how to define geometric quantities
that generalize connection and curvature,
when the metric is replaced by the area metric.
We will refer to these quantities as
area connection and area curvature,
or just connection and curvature
when there is no risk of confusion.

\section{Area connection}
\label{area-connection}

\subsection{Transformation of area connection}
\label{transf-conn}

The notion of area (\ref{da=h}) can be interpreted as the norm 
on a fiber in the bundle of 2-forms on the space-time manifold.
The connection on the bundle is a 1-form
\be
\Gb^{[ij]}{}_{[kl]} = 
dx^{m}\Gb_m{}^{[ij]}{}_{[kl]},
\ee
and we call it the {\em area connection}.
Here a pair of anti-symmetrized indices $[ij]$ 
is used to label the basis $dx^{i}\wedge dx^{j}$ 
of the space of 2-forms.

The area connection $\Gb^{ij}{}_{kl}$ is defined so that
the covariant derivative of a rank-2 anti-symmetric tensor field
\be
(DV)^{ij} \equiv
dV^{ij} + \frac{1}{2} \Gb^{ij}{}_{kl} V^{kl}
\label{DV-upper}
\ee  
is covariant.
Under a general coordinate transformation
$x^i \rightarrow x'{}^i(x)$,
since a tensor field $V^{ij}$ transforms as
\be
V^{ij} \quad \rightarrow \quad
V^{\prime ij} = M^i{}_k M^j{}_l V^{kl},
\ee
where
\be
M^i{}_j \equiv \frac{\del x^{\prime i}}{\del x^j},
\ee
the area connection should transform as
\be
\Gb^{\prime ij}{}_{kl}
= M^i{}_p M^j{}_q \Gb^{pq}{}_{rs} M^{-1 r}{}_k M^{-1 s}{}_l
- 2 d(M^{[i}{}_r M^{j]}{}_s) M^{-1 r}{}_{[k} M^{-1 s}{}_{l]},
\label{transf-Gamma}
\ee
so that
the covariant derivative transforms covariantly as
\be
(DV)^{\prime ij} = M^i{}_k M^j{}_l (DV)^{kl}.
\ee

Since the covariant derivative of a scalar is just 
the ordinary derivative,
in order to preserve the Leibniz rule 
\be
d(V^{ij}W_{ij}) \equiv
D(V^{ij}W_{ij}) = (DV)^{ij}W_{ij} + V^{ij}(DW)_{ij},
\ee
the covariant derivative of the tensor 
with lower indices should be defined as
\bea
DW_{ij} &=& 
dW_{ij} - \frac{1}{2} W_{kl} \Gb^{kl}{}_{ij}.
\label{DV-lower}
\eea
The Leibniz rule can be used to determine
how the covariant derivative acts on a tensor
with an arbitrary set of pairs of anti-symmetrized indices.

According to the transformation law 
of the area connection (\ref{transf-Gamma}), 
the combination
\be
\Tb_{[k}{}^{[ij]}{}_{lm]} \equiv
\Gb_k{}^{[ij]}{}_{[lm]}
+\Gb_l{}^{[ij]}{}_{[mk]}
+\Gb_m{}^{[ij]}{}_{[kl]}
\label{T}
\ee
is covariant,
and can be viewed as the generalization 
of the torsion defined for an ordinary connection \cite{Schuller:2005yt}.
We refer to this quantity as the {\em area torsion}.

Naively,
the fact that, 
apart from its 1-form index,
the indices of the area connection $\Gb$ are anti-symmetric pairs
suggests that covariant derivatives apply to tensors 
only if all indices of the tensor are anti-symmetric pairs,
and thus only to tensors with an even number of indices.
This is not entirely true.
For an arbitrary tensor of even or odd rank,
we can define a differential form from it by 
contracting some of its indices with $dx^i$'s.
The covariant derivative can then apply to this differential form 
as long as the remaining indices of the differential form
come in anti-symmetric pairs.
For example, 
the torsion $\Tb_i{}^{jk}{}_{lm}$ is a tensor of rank 5.
We can define a 1-form
\be
\Tb^{ij}{}_{kl} \equiv dx^m \; \Tb_m{}^{ij}{}_{kl}
\ee
and the covariant derivative on the torsion can be defined as
\be
(D\Tb)^{ij}{}_{kl} \equiv \frac{1}{2} dx^m\wedge dx^n
\left[\left(
\del_m\Tb_n{}^{ij}{}_{kl} 
+ \Gb_m{}^{ij}{}_{pq}\Tb_n{}^{pq}{}_{kl}
- \Gb_m{}^{pq}{}_{kl}\Tb_n{}^{ij}{}_{pq}
\right) - (m\leftrightarrow n)\right].
\ee

\subsection{Area metricity and area torsion-free conditions}
\label{compatible-metric}

In Riemannian geometry, 
the Levi-Civita connection is uniquely fixed by the metricity condition
and the torsion-free condition,
and the Einstein equation and Hilbert-Einstein action
can be expressed in terms of its curvature.
Similarly, 
we would like to impose various constraints on the area connection $\Gb^{ij}{}_{kl}$,
to look for the counterpart of the Levi-Civita connection for the area metric.

In view of its importance in the formulation of physical laws,
it is natural to impose the area metricity condition:
\be
(Dh)_{ijkl} = 0,
\label{Dh=0}
\ee
where
\be
(Dh)_{[ij][kl]} \equiv
dh_{[ij][kl]} - \frac{1}{2} h_{[mn][kl]} \Gb^{[mn]}{}_{[ij]}
- \frac{1}{2} h_{[ij][mn]} \Gb^{[mn]}{}_{[kl]}.
\label{Dh-lower}
\ee
As a result,
the inverse $h^{ijkl}$ is also covariantly constant.
Using $h_{ijkl}$ and $h^{ijkl}$,
one can lower or raise a pair of anti-symmetrized indices.
For example,
\be
V_{ij} \equiv \frac{1}{2}h_{ijkl} V^{kl}.
\ee

The area metricity condition implies that the area torsion
vanishes for the totally anti-symmetrized combination,
\be
\Tb_{[ijklm]} \equiv h_{pq[jk} \Tb_{i}{}^{pq}{}_{lm]} = 0,
\label{T5=0}
\ee
where all 5 indices $\{i, j, k, l, m\}$ are totally anti-symmetrized.

In analogy with Riemannian geometry,
it is desirable for the area connection 
to be torsion-free,
that is,
\be
{\Tb}_{[k}{}^{[ij]}{}_{lm]} = 0.
\label{T=0}
\ee

In the context of quantum gravity
when the background is a superposition of
coherent states of different metrics,
the area metric $h$ is defined directly
in terms of a set of metrics $\{g^{(\a)}\}$ through (\ref{h=gg-A}).
It would then be natural to define \cite{Schuller:2005yt}
\be
\Gb_{ikljm} \equiv h_{klpq}\Gb_{i}{}^{pq}{}_{jm}
=\sum_{\a} \left(
[\G^{(\a)}_{ikj}g^{(\a)}_{lm} - (k\leftrightarrow l)]-[(j \leftrightarrow m)]
\right)
\label{Gamma-gg}
\ee
where $\G^{(\a)}$ is a connection for the metric $g^{(\a)}$,
and this area connection automatically satisfies the area metricity condition (\ref{Dh=0}).
If $\G^{(\a)}$ is given by the Christoffel symbol
(so that it is torsion-free),
this expression also leads to
the vanishing of the torsion (\ref{T=0}).
Note, however, 
that there are infinitely many sets of metrics $\{g^{(\a)}\}$
corresponding to the same area metric.
Hence there are infinitely many choices of
the area connection $\Gb$ that satisfy both
the area metricity (\ref{Dh=0})
and the torsion-free conditions (\ref{T=0}).

In the context of string theory,
when $h$ is the only quantity that defines the background geometry,
the choice of area connection through a set of metrics $\{g^{(\a)}\}$ (\ref{Gamma-gg}) is
not appropriate because it is not independent of the choice of 
the set of metrics. 

Another way to see that there are infinitely many solutions 
to both the area metricity condition and the area torsion-free condition 
is to check that these two conditions are satisfied by
the expression of the area connection as
\be
\Gb_{mijkl} = \frac{1}{2} \del_m h_{ijkl} 
+ \frac{1}{4}( \del_i h_{jmkl} - \del_j h_{imkl} - \del_k h_{ijlm} + \del_l h_{ijkm} ).
\ee
One can check that this equation is not covariant.
That is, the transformation of the area metric leads to 
a transformation law of the right hand side of the equation 
that is not compatible with the transformation law of the area connection.
This means that we can use this expression in 
different coordinate systems and obtain different area connections 
compatible with area metricity and are area torsion-free.

\subsection{Induced connection}
\label{InducedConnection}

The transformation law of the area connection (\ref{transf-Gamma})
allows us to define the connection 1-form
\cite{Schuller:2005yt}
\be
\G^i{}_j(0) \equiv \frac{1}{D-2} \Gb^{ik}{}_{jk} 
- \frac{1}{2(D-1)(D-2)}\d^i_j\Gb^{kl}{}_{kl},
\label{def-G-1}
\ee
which transforms like an ordinary connection in Riemannian geometry
\be
\G^i{}_j(0) \rightarrow
\G^{\prime i}{}_j(0) = 
M^i{}_k \G^{k}{}_{l}(0) M^{-1 l}{}_j
- (dM^{i}{}_k) M^{-1 k}{}_j.
\ee

The reason for the notation ``$(0)$'' is that
this is not the only combination of components of $\Gb$ 
that transforms like an ordinary connection.
There are infinitely many of them 
and we will call them {\em induced connections}.
We can readily write down many other induced connections as
\be
\G_j{}^i{}_k(\lam) = \G_j{}^i{}_k(0) - \lam T_j{}^i{}_k,
\label{Gijklam}
\ee
where $T_j{}^i{}_k$ is the torsion of $\G_j{}^i{}_k(0)$:
\bea
T_j{}^i{}_k &\equiv& \G_j{}^i{}_k(0) - \G_k{}^i{}_j(0).
\label{torsion-2}
\eea
The induced connection 
$\Gamma_j{}^i{}_k(\lam = 1/2)$ is torsion-free.

The induced connections $\G_j{}^i{}_k(\lam)$ (\ref{Gijklam}) allow us to 
define covariant derivatives that can act on indices individually,
not only on pairs of anti-symmetrized indices.
In particular,
the volume form (\ref{Omega}) is covariantly constant with respect to $\G{}^i{}_j(0)$.
That is,
\be
d\om\eps_{i_1\cdots i_D} = 
\om\eps_{ji_2\cdots i_D}\G^j{}_{i_1}(0)
+ \cdots 
+ \om\eps_{i_1\cdots i_{D-1}j}\G^j{}_{i_D}(0).
\label{Domega=0}
\ee
It is equivalent to the relation
\be
\G^j{}_j(0) = \om^{-1} d\om,
\ee
which is guaranteed by the area metricity condition.
In this sense,
the induced connection $\G^i{}_j(0)$ with $\lam = 0$ is special.

The volume form is covariantly constant with respect to 
other induced connections $\G^i{}_j(\lam)$ with $\lam \neq 0$
only if $T_i{}^j{}_j = 0$.
Since
\be
T_i{}^j{}_j = - \frac{1}{2(D-2)} {\Tb}_i{}^{jk}{}_{jk},
\label{TT}
\ee
the volume form is covariantly constant 
with respect to all $\G^i{}_j(\lam)$
if the area torsion vanishes.

For a given induced connection $\G^i{}_j(\lam)$,
one can decompose $\Gb^{ij}{}_{kl}$ as
\be
\Gb^{ij}{}_{kl} \equiv
\Sb^{ij}{}_{kl}(\lam) + \hat{\bf\G}^{ij}{}_{kl}(\lam),
\label{decompose-G}
\ee
where $\Sb^{ij}{}_{kl}(\lam)$ is a covariant tensor,
and
\be
\hat{\bf\G}^{ij}{}_{kl}(\lam)
\equiv
\G^i{}_k(\lam)\d^j_l - \G^i{}_l(\lam)\d^j_k
- \G^j{}_k(\lam)\d^i_l + \G^j{}_l(\lam)\d^i_k.
\label{hat-G}
\ee
The tensor $\Sb(0)$ is traceless:
\be
\Sb^{ik}{}_{jk}(0) = 0.
\label{trS=0}
\ee
The difference between $\Sb_m{}^{ij}{}_{kl}(\lam)$
and $\Sb_m{}^{ij}{}_{kl}(0)$
is $\lam$ times 
\be
\left[
T_m{}^i{}_k\d^j_l - T_m{}^i{}_l\d^j_k
- T_m{}^j{}_k\d^i_l + T_m{}^j{}_l\d^i_k
\right].
\ee

In Riemannian geometry,
torsion is the only covariant quantity one can derive
algebraically (without taking derivatives) 
from a connection $\G^i{}_j$.
In other words,
it is the only tensor one needs to fix in order to uniquely determine 
the connection for a given metric.
For the area connection $\Gb$,
we have not only the area torsion $\Tb$ as a covariant tensor
defined algebraically from a given area connection $\Gb$,
but also the torsion $T_i{}^j{}_k$ for the induced connection
and the tensor $\Sb^{ij}{}_{kl}(\lam)$ (for any fixed value of $\lam$).
To uniquely determine an area connection for a given area metric $h_{ijkl}$,
one needs to fix all these covariant quantities.

\subsection{Parametrization of area metric and area connection}

In 2D, 
there is a single component $h_{1212}$ in the area metric.
At any given point, 
it can be scaled to $\pm 1$ by coordinate transformation.

In 3D,
we can always find a metric $g_{ij}$ (up to sign) to represent $h_{ijkl}$.
The difference between metric and area metric is less significant.

In 4D,
there are 20 independent components in $h$ 
satisfying (\ref{h-symm-1}) and (\ref{h-symm-2}).
Roughly speaking,
using the 16 independent parameters in a $GL(4)$ transformation,
we should be able to use $4$ parameters to parametrize $h$ 
at a given point. 
But there are more canonical forms of the area metric, 
categorized into 23 meta-classes \cite{Schuller:2009hn}.
In contrast, 
a metric $g_{ij}$ can always be cast in the canonical form
$g_{ij} = \mbox{diag}(\pm 1, \pm 1, \pm 1, \pm 1)$
at any point by coordinate transformation.

For the Taylor expansions of $h_{ijkl}$ around a given point $x_0$ in 4D
\be
h_{ijkl}(x) = h_{ijkl}(x_0) + (x-x_0)^m \del_m h_{ijkl}(x_0) + \cdots,
\ee
there are $4\times 20$ independent coefficients $\del_m h_{ijkl}(x_0)$
at the first order.
On the other hand,
there are $4\times 10$ coefficients $\del_j\del_k x^{i\prime}(x_0)$
at the second order in the Taylor expansion of 
the new coordinates $x^{i\prime}(x)$
\be
x^{i\prime}(x) = x^i + (x-x_0)^j \del_j x^{i\prime}(x_0)
+ \frac{1}{2} (x-x_0)^j (x-x_0)^k \del_j\del_k x^{i\prime}(x_0)
+ \cdots
\ee
as a function of the old coordinates $x^j$.
The latter coefficients change the former under a coordinate transformation.
Thus we expect to be able to construct out of $(\del_m h_{ijkl})(x_0)$
$80-40=40$ covariant quantities.
These quantities are encoded in the tensor $S^{ij}{}_{kl}(\lam)$.

This is in contrast with the case of the metric $g_{ij}$.
In $D$ dimensions,
there are $D\times \frac{D(D+1)}{2}$ coefficients $\del_k g_{ij}(x_0)$
at the linear order in the expansion of $g_{ij}$ around a point $x_0$, 
and as many coefficients $\del_j \del_k x^{i\prime}(x_0)$ 
in the expansion of $x^{i\prime}$ at the quadratic order,
so covariant geometric quantities (the curvature tensor)
are only encoded in second or higher order terms 
in the expansion of $g_{ij}$
in Riemann normal coordinates.

\subsection{Generalized geodesic equation}
\label{GGE}

Another condition one may wish to impose
on the area connection is its compatibility with
the generalized geodesic equation,
namely the equation of motion for a string
to extremize its world-volume.
In this subsection,
we derive the generalized geodesic equation
and find its compatibility condition with the area connection.

In terms of the bracket
\be
\{f, g\} \equiv \frac{\eps^{ab}(\del_{a}f)(\del_{b}g)}
{{\cal A}_h},
\ee
where 
${\cal A}_h$ is defined in (\ref{H}),
the equation of motion of the generalized Nambu-Goto action
(\ref{generalized-NG}) is
\be
h_{ijkl}\{X^{j}, \{X^{k}, X^{l}\}\}
+ \Gb^{NG}_{i([jk][lm])}\{X^{j}, X^{k}\}\{X^{l}, X^{m}\}
= 0,
\label{generalized-geodesic}
\ee
which is analogous to the geodesic equation.
\footnote{
The geodesic equation can be written as
\be
g_{ij}\frac{d^2 x^{j}}{d s^2} 
+ (g\Gamma)_{ijk}\frac{dx^{j}}{ds}\frac{dx^{k}}{ds} 
= 0 
\qquad
\mbox{with}
\quad
(g\Gamma)_{ijk} = 
\frac{1}{2}
\left(
\del_{j}g_{ik} + \del_{k}g_{ij} - \del_{i}g_{jk}
\right),
\ee
where the derivative $d/ds$ can be viewed as 
the generator of length-preserving diffeomorphism, 
while $\{\cdot, \cdot\}$ in (\ref{generalized-geodesic})
generates area-preserving diffeomorphism.
}
Here $\Gb^{NG}$ is defined by
\be
\Gb^{NG}_{i([jk][lm])}
\equiv \frac{1}{4}\left(
\del_{i}h_{jklm}
-\del_{j}h_{iklm}
+\del_{k}h_{ijlm}
+\del_{l}h_{mijk}
-\del_{m}h_{lijk}
\right).
\label{Gamma-dh}
\ee
The indices in $[\cdots]$ are anti-symmetrized, 
and the two pairs of indices in $(\cdots)$ are symmetrized.

Recall that the ordinary geodesic equation is equivalent 
to the covariant constancy of 
the tangent vector $dx^i/ds$
of the geodesic curve in the direction of $dx^i/ds$.
More precisely,
for a vector field $V^j(x)$,
we define the covariant derivative in the direction of $dx^i/ds$ as
\be
\frac{dx^j}{ds}\left(D_j V^i\right)_{x\rightarrow x(s)} 
= \frac{dx^j}{ds} \left(\frac{\del}{\del x^j}V^i + \Gamma_j{}^i{}_k V^k\right)_{x\rightarrow x(s)} 
= \frac{d}{ds}V^i(x(s))  + \frac{dx^j}{ds}\Gamma_j{}^i{}_k V^k(x(s)) \equiv \frac{D}{ds} V^j(x(s)),
\label{dD}
\ee
so that the geodesic equation can be written as
\be
\frac{D}{ds}\frac{dx^i}{ds} = 0.
\ee
Note that we have replaced the vector field $V^j(x)$ in (\ref{dD})
by $dx^j/ds$, 
which is a function of the world-line parameter $s$.

We will say that 
the generalized geodesic equation (\ref{generalized-geodesic})
is compatible with an area connection $\Gb$ if
it can be derived from the covariant constancy condition 
for $\{X^i, X^j\}$
in the ``direction'' of $\{X^i, X^j\}$.
Naturally,
for an anti-symmetric tensor field $W^{lm}(X)$,
it is defined as
\bea
\{X^j, X^k\} \left(D_j W_{ki}\right)_{X\rightarrow X(\s)}
&=& \{X^j, X^k\} h_{kilm} \left(D_j W^{lm}\right)_{X\rightarrow X(\s)}
\nn \\
&=& \{X^j, X^k\} h_{kilm}\left(\del_j W^{lm} + \Gb_j{}^{lm}{}_{pq} W^{pq}\right)_{X\rightarrow X(\s)}
\nn \\
&=& - h_{kilm} \left[ \{X^k, W^{lm}(X(\s))\} - \{X^j, X^k\} \Gb_j{}^{lm}{}_{pq} W^{pq}(X(\s)) \right]
\nn \\
&=& h_{iklm} \{X^k, \{X^l, X^m\}\} + \Gb_{jkipq} \{X^j, X^k\} W^{pq}.
\eea
The condition for $\{X^i, X^j\}$ to be covariantly constant in
the ``direction'' of $\{X^i, X^j\}$ is thus
\be
h_{ijkl} \{X^j, \{X^k, X^l\}\} + \Gb_{jkilm} \{X^j, X^k\} \{X^l, X^m\} = 0
\ee
Comparing this equation with 
the generalized geodesic equation (\ref{generalized-geodesic}),
we see that they can be identified with each other if
$\Gb^{NG}$ can be identified with $\Gb$.
Assuming the area metricity condition (\ref{Dh=0}),
we can rewrite the first derivatives of the area metric in (\ref{Gamma-dh})
in terms of the area connection.
Under the contraction with $\{X^j, X^k\} \{X^l, X^m\}$,
$\Gb^{NG}$ is equivalent to 
\be
\Gb^{NG}_{i([jk][lm])}
\sim 
\Gb_{jkilm} + \frac{1}{2} \Tb_{ijklm},
\label{GNG-G}
\ee
without assuming the area torsion-free condition.
In other words,
if the area torsion $\Tb$ vanishes,
the generalized geodesic equation (\ref{generalized-geodesic})
can be interpreted as the condition for the tangent plane of the string
to be parallel transported along its world-sheet.

Strictly speaking,
it is unnecessary for the area torsion to vanish 
for the compatibility of the area connection with
the generalized geodesic equation,
one only needs the contraction of the area torsion $\Tb_{ijklm}$ 
with the factor $\{X^j, X^k\}\{X^l, X^m\}$ to vanish.
For instance,
since
\bea
\{X^j, X^k\}\{X^l, X^m\} 
&=&
\frac{1}{{\cal A}_h^2}
(\del_{\a}X^j\del_{\b}X^k\del^{\a}X^l\del^{\b}X^m 
- \del_{\a}X^j\del_{\b}X^k\del^{\b}X^l\del^{\a}X^m),
\eea
where the first term symmetrizes $(j, l)$ and $(k, m)$,
and the second term can be obtained from the first term 
by anti-symmetrizing $(l, m)$,
according to (\ref{GNG-G}),
$\Gb^{NG} \sim \Gb$ if
\be
[(\Tb_{i[jk]lm} + \Tb_{i[lm]jk}) + (j \leftrightarrow l)]
+ [(k \leftrightarrow m)] = 0,
\label{TTTT}
\ee
which does not require that the area torsion vanishes identically.

\section{Torsion-free area connections}
\label{more-connection}

\subsection{Construction of torsion-free area connections}

In Riemannian geometry,
given any connection $\G$, 
we can always derive from it a torsion-free connection 
$\G'_{j}{}^{i}{}_{k} \equiv \G_{j}{}^{i}_{k} - \frac{1}{2}T_{j}{}^{i}_{k}$,
which satisfies the same metricity condition as $\G$.
(Here $T_{j}{}^{i}_{k}$ is the torsion of the connection $\G$.)
We will show in this section that 
similarly one can add a tensor to any area connection 
such that the new area connection is torsion-free (\ref{T=0}),
and satisfies the same area metricity condition (\ref{Dh=0}) and (\ref{Dh-lower}),
although the algebra involved is more complicated.

One can add a tensor $\Kb^A{}_B$ 
to any area connection such that
the new area connection 
\be
\Gb'^A{}_B \equiv \Gb^A{}_B + \Kb^A{}_B
\qquad 
(A = [ij], B = [kl])
\label{Gp}
\ee
preserves the area metricity condition
by demanding $\Kb$ to be anti-symmetric
\be
\Kb_{AB} + \Kb_{BA} = 0.
\label{K-K=0}
\ee

We aim at introducing a minimal amount of freedom in $\Kb$
just enough to realize the area torsion-free condition
for the new area connection.
A simple possibility is
\be
\Kb_{ijklm} \equiv \Ub_{ijklm} - \Ub_{ilmjk},
\label{K=U-U}
\ee
where the free tensor $\Ub$ is required to 
have the same symmetry as the area torsion,
\bea
&\Ub_{ijklm} = \Ub_{ljkmi} = \Ub_{mjkil},
\label{U1}
\\
&\Ub_{ijklm} = - \Ub_{ikjlm},
\qquad 
\Ub_{ijklm} = - \Ub_{ijkml},
\label{U2}
\\
&\Ub_{[ijklm]} = 0,
\label{U3}
\eea
so that it has the same number of independent components
as the area torsion.

More generally,
we can use the ansatz
\be
\Kb_{ijklm} \equiv \Ub_{ijklm} - \Ub_{ilmjk}
+ \a (\Ub_{[jk]ilm} - \Ub_{[lm]ijk}),
\label{Kb-general}
\ee
with a free parameter $\a$ 
as a generalization of (\ref{K=U-U}).

The condition of zero area torsion is a set of linear relations for $\Ub$.
Due to the symmetries (\ref{U1})--(\ref{U3}) of the tensor $\Ub$,
these relations are coupled to each other in different patterns 
depending on how many different values the five indices $(i,j,k,l,m)$
of the new torsion $\Tb'_{ijklm}$ take.
When they take only 3 different values, say, $(1, 2, 3)$,
the condition for the new torsion $\Tb'_{1[12]23}$ to vanish is
\be
-\Tb_{1[12]23} = 2(1-\alpha)\Ub_{1[12]23}.
\ee
$\Ub_{1[12]23}$ can be easily determined 
as long as $\alpha \neq 1$.

When there are four different values among the five indices of $\Tb'_{ijklm}$,
we have a set of linear equations to solve:
\bea
-\Tb_{1[34]23} &=&
(3-\a/2)\Ub_{1[34]23} - (1+3\a/2)\Ub_{1[23]34} + (1+3\a/2)\Ub_{2[13]34},
\\
-\Tb_{2[13]34} &=&
(3-\a/2)\Ub_{2[13]34} - (1+\a/2)\Ub_{1[34]23} - (1+3\a/2)\Ub_{1[23]34},
\\
-\Tb_{1[23]34} &=&
(3-\a/2)\Ub_{1[23]34} - (1+3\a/2)\Ub_{1[34]23} - (1+3\a/2)\Ub_{2[13]34}.
\eea
These unknowns of $\Ub$ can be uniquely solved if 
the determinant of the $3\times 3$ matrix of coefficients 
in these linear relations is non-zero.
The determinant is $(1-\a)^2(1+\a/2)$.
Hence we need $\a \neq 1$ and $\a \neq -2$.

When all five indices of $\Tb'_{ijklm}$ take different values,
one needs to solve 10 linear relations for 10 variables, say,
$\Ub_{12345}, \Ub_{14523}, \Ub_{23145}, \Ub_{32145}, \Ub_{45123},
\Ub_{54123}, \Ub_{23451}, \Ub_{45231}, \Ub_{32451}$ and $\Ub_{23541}$.
There are in fact only 9 independent variables because
of the constraint (\ref{U3}).
Correspondingly the torsion $\Tb_{ijklm}$ satisfies the same identity (\ref{T5=0})
and the 10 linear relations are linearly dependent.
Thus the task is to solve a reduced set of 9 linear relations for 9 independent variables.
The determinant of the $9\times 9$ matrix of coefficients is
$(1-\a)^5 (1+\a/2)^4$.
It also demands that $\a \neq 1, -2$.

In 3 dimensions,
there are at most 3 different values for the indices to take,
and the coefficient $\a$ is redundant.

In 4 or higher dimensions,
there are $\Tb'$ components with indices taking 4 or 5 different values.
The tensor $\Ub$ can be uniquely fixed by 
the area torsion-free condition (\ref{T=0}) 
for the new area connection $\Gb'$
as long as
\be
\a \neq 1, 
\quad
\mbox{and}
\quad
\a \neq -2.
\ee

\subsection{Generalized Levi-Civita connection} 
\label{GLCC}

In the discussions on physical theories of gravity
(the dynamics of space-time geometry),
the choice of a connection is in some sense a convention.
If a connection is used in a gravitational theory 
to write down its field equation,
one can always rewrite it in terms of another connection,
merely by a field redefinition.
For example, 
the Levi-Civita connection is conventionally adopted
in Einstein's theory.
But teleparallel gravity,
which is equivalent to Einstein's theory,
is formulated in terms of the Weitzenb\"{o}ck connection,
which has zero curvature but non-zero torsion.
The choice of connection depends on 
the choice of formulation of the physical theory of gravity.

Nevertheless, 
the Levi-Civita connection has its conceptual advantage.
With torsion viewed as the covariant information contained in a connection,
the Levi-Civita connection is the special connection 
for which this information is empty.
It can be used as a reference connection.

In this section, 
we aim at defining the counterpart of Levi-Civita connection 
for the geometry of area metric.
This will provide a reference area connection 
for the convenience of future discussions.
More importantly,
this task 
helps us understand better the content of the area connection.
This understanding will be useful 
when we construct a dynamical theory for the area metric.

As a generalization of the Levi-Civita connection,
the reference area connection should satisfy
both the area metricity condition (\ref{Dh=0})
and the area torsion-free condition (\ref{T=0}).
It was commented in Sec. \ref{compatible-metric} that 
additional constraints are needed to fix the area connection uniquely.
The problem is to find suitable constraints just enough
to fix the area connection without inconsistencies 
with the area metricity and the area torsion-free conditions.

The basic idea is to introduce just enough degrees of freedom in a simple way
in the area connection to satisfy the constraints,
i.e., the area metricity condition 
and the area torsion-free condition,
so that the constraints can fix the area connection uniquely.

In view of the previous subsection,
where we showed how to turn any area connection into one
that is area torsion-free,
all we need is to find an area connection that satisfies the area metricity condition,
and then we can modify it by adding the tensor $\Kb$.
There are, of course,
infinitely many area connections satisfying the area metricity condition.
Our task is to find one such connection that is simple,
and can be easily related to the Levi-Civita connection when 
the area metric is defined from a Riemannian metric through (\ref{h=gg}).

First,
we know that the area connection needs to include a part
(\ref{hat-G}) 
\be
\hat{\Gb}_k{}^{ij}{}_{lm}(\lam) \equiv 
\Gamma_k{}^i{}_l(\lam) \d^j_m - \Gamma_k{}^j{}_l(\lam) \d^i_m
- \Gamma_k{}^i{}_m(\lam) \d^j_l + \Gamma_k{}^j{}_m(\lam) \d^i_l,
\label{Gbhat}
\ee
which is composed of the induced connection
in order for it to transform properly,
where $\Gamma_j{}^i{}_k(\lam)$ is defined by
(\ref{def-G-1}) and (\ref{Gijklam}).
While $\hat{\Gb}$ may not satisfy the area metricity condition,
we need to add more degrees of freedom in $\Gb$.

The area metricity condition is a tensorial equation of the form 
\be
(\mbox{Area Metricity})_{iAB} = 0,
\ee
with the indices $A, B$ symmetrized.
Correspondingly,
one can introduce a tensor $\Sb'$ with the same symmetry
\be
\Sb'_{iAB} = \Sb'_{iBA}
\label{Sbp-symm}
\ee
into the area connection.
But we need to impose an additional constraint on $\Sb'$
to remove as many degrees of freedom in $\Sb'$
as there are in the induced connection $\G_i{}^j{}_k(\lam)$.
For $\lam = 0$,
it is the traceless condition
\be
\Sb'_i{}^{jl}{}_{kl}(0) = 0.
\label{Sbp-traceless}
\ee

Therefore, generically, 
the sum $\Gb = \Sb'(\lam) + \hat{\Gb}(\lam)$ is expected to be uniquely determined 
by the area metricity condition.
For a given area metric,
this expression depends on $\lam$,
since the choice of the parameter $\lam$ affects 
the meaning of the constraint (\ref{Sbp-symm}).

The next step is to add $\Kb$ to the area connection 
as we described in the previous subsection.

Instead of imposing the area metricity condition 
and then the area torsion-free condition in two steps,
one can simply put all these degrees of freedom
$\Sb'$, $\hat{\Gb}$ and $\Ub$
together to construct an area connection as
\be
\Gb 
= \Sb'(\lam) + \Kb(\lam) + \hat{\Gb}(\lam),
\label{Gb-def1}
\ee
where 
$\Sb'(\lam) = \Sb'(0) + \lam \hat{\Tb}$,
$\hat{\Gb}(\lam) = \hat{\Gb}(0)  - \lam \hat{\Tb}$
and
\be
\hat{\Tb}_k{}^{ij}{}_{lm} \equiv
T_k{}^i{}_l \d^j_m - T_k{}^j{}_l \d^i_m
- T_k{}^i{}_m \d^j_l + T_k{}^j{}_m \d^i_l,
\ee
with $T_j{}^i{}_k$ being the torsion of the induced connection (\ref{torsion-2})
\be
T_j{}^i{}_k \equiv \Gamma_j{}^i{}_k(0) - \Gamma_k{}^i{}_j(0).
\ee
The tensor $\Sb'$ is subject to
the conditions (\ref{Sbp-symm}) and (\ref{Sbp-traceless}),
which now become
\bea
&\Sb'_{iAB}(\lam) = \Sb'_{iBA}(\lam).
\label{Sbp-symm-2}
\\
&(\Sb'(\lam) - \lam \hat{\Tb})_i{}^{jl}{}_{kl} = 0,
\label{Sbp-traceless-2}
\eea

A bunch of modifications of the constraints on $\Sb'(\lam)$ and $\Kb(\lam)$ are possible,
without changing the number of independent degrees of freedom 
in the area connection, 
so that it will still be uniquely fixed by the area metricity condition 
and the area torsion-free condition.
For instance,
the condition (\ref{Sbp-traceless-2}) can be modified as
\be
(\Sb'(\lam) + (1-\beta) \Kb(\lam) - \lam \hat{\Tb})_i{}^{jl}{}_{kl} = 0.
\label{Sbp-traceless-3}
\ee
This corresponds to the constraint (\ref{trS=0})
after replacing $\Gb$ by $\Gb - \beta \Kb$ 
in (\ref{def-G-1}) to define the induced connection.

To summarize, 
the area connections as candidates of reference area connections
are a sum of three terms as (\ref{Gb-def1}),
where $\Sb'(\lam)$ is constrained by (\ref{Sbp-symm-2}) and (\ref{Sbp-traceless-3}),
$\hat{\Gb}(\lam)$ is defined by (\ref{def-G-1}) and (\ref{Gbhat}),
and $\Kb(\lam)$ is given by (\ref{Kb-general})
with $\Ub$ satisfying (\ref{U1})--(\ref{U3}).

There are thus three parameters $\lam, \alpha$ and $\beta$
to parametrize the area connection $\Gb(\lam, \a, \b)$.
The simplest choice of the parameters $\lam, \alpha$ and $\beta$ is
\bea
\lam = 0, 
\qquad 
\alpha = 0,
\qquad 
\beta = 1,
\eea
and we will refer to the corresponding area connection 
as the area Levi-Civita connection.
For this case,
the area connection is the sum of three terms
\be
\Gb_{mijkl} = \Sb'_{mijkl} + \Kb_{mijkl} + \hat{\Gb}'_{mijkl},
\label{Gb-expansion-2}
\ee
where $\Sb'$ is subject to 
the constraints
\bea
&\Sb'_i{}^{jl}{}_{kl} = 0,
\label{Sb-1}
\\
&\Sb'_{iAB} = \Sb'_{iBA},
\label{Sb-2}
\eea
$\Kb$ is defined as
\bea
&\Kb_{mijkl} = \Ub_{mijkl} - \Ub_{mklij},
\label{Kb-1}
\eea
where $\Ub$ satisfies (\ref{U1})--(\ref{U3}),
which is reproduced here
\bea
&\Ub_{ijklm} = \Ub_{ljkmi} = \Ub_{mjkil},
\label{U1-2}
\\
&\Ub_{ijklm} = - \Ub_{ikjlm},
\qquad
\Ub_{ijklm} = - \Ub_{ijkml},
\label{U2-2}
\\
&\Ub_{[ijklm]} = 0,
\label{U3-2}
\eea
and $\hat{\Gb}'$ is defined as 
\be
\hat{\Gb}'_k{}^{ij}{}_{lm} \equiv 
\Gamma'_k{}^i{}_l \d^j_m - \Gamma'_k{}^j{}_l \d^i_m
- \Gamma'_k{}^i{}_m \d^j_l + \Gamma'_k{}^j{}_m \d^i_l.
\label{Gbhat-2}
\ee
The connection $\Gamma'_j{}^i{}_k$ is different from
$\Gamma_j{}^i{}_k$ defined by (\ref{def-G-1}).
Instead it satisfies a relation of the form (\ref{def-G-1}),
but with $\Gb$ replaced by $\Gb - \Kb$.
Since this relation is a direct consequence of 
(\ref{Gb-expansion-2}) and (\ref{Sb-1}),
it does not have to be imposed separately.

The calculation of the area Levi-Civita connection 
for a given area metric
can be proceeded as follows.

\begin{enumerate}

\item
Solve the area metricity condition (\ref{Dh=0}),
which is equivalent to
\be
\del_i h_{[jk][lm]} = 
2\Sb'_{i[jk][lm]} + \hat{\Gb}'_{i[jk][lm]} 
+ \hat{\Gb}'_{i[lm][jk]}.
\label{dh-2}
\ee
Contracting two indices of this equation, say $k$ and $m$
(with the help of $h^{-1}$),
the tracelessness of $\Sb'$ (\ref{Sb-1}) 
allows us to solve $\hat{\Gb}'$
(i.e. to solve $\Gamma'_j{}^i{}_k$),
unless there are degeneracies in the area metric.
\footnote{
The degenerate cases will be discussed below through an example.
}

\item
After obtaining $\G'_j{}^i{}_k$,
(\ref{dh-2}) can be used to determine $\Sb'$.

\item
Finally,
$\Kb$ can be calculated as described 
in the previous subsection 
by demanding the area torsion to vanish.

\end{enumerate}

To make connection with Riemannian geometry in which 
the Levi-Civita connection is used, 
when $h$ is defined by a metric $g$ (\ref{h=gg}),
it is easy to see that all conditions for 
the area Levi-Civita connection is satisfied by
setting $\Sb' = \Kb = 0$ and
$\G'_j{}^i{}_k$ equal to the Christoffel symbol.

We have no intention to claim that 
this is the only way to uniquely determine the area connection 
for a given area metric.
The simple fact is that 
in the case of area geometry,
there are a lot more covariant tensorial degrees of freedom 
in the area connection 
(while in Riemannian geometry
the torsion is the only one),
and as a result
there are a lot of different ways to establish 
a connection between the area metric 
and the area connection.

In the context of string theory,
the requirement of conformal symmetry should 
impose an equation on the area connection.
A good choice of the area connection would be the one
in terms of which that equation can be expressed in a simple form.
However, 
this is out of the scope of this work.
We leave this problem for future study.

\section{Area curvature}
\label{area-curvature}

\subsection{Area curvature and induced curvature}

The area curvature 2-form can be defined for an area connection $\Gb^{ij}{}_{kl}$ as
\be
{\Rb} = (d + \Gb)^2,
\label{R}
\ee
or more explicitly,
\be
{\Rb}^A{}_B \equiv
\frac{1}{2} \Rb_{ij}{}^A{}_B dx^i \wedge dx^j
= d\Gb^A{}_B + \Gb^A{}_C \wedge \Gb^C{}_B,
\ee
where $A, B, C$ represent pairs of anti-symmetrized indices.
The area curvature tensor ${\Rb}_{ij}{}^{kl}{}_{mn}$ 
has 3 pairs of anti-symmetrize indices.

It is straightforward to check that
\be
{\Rb}_{ijklmn} = - {\Rb}_{ijmnkl},
\ee
where 
\be
{\Rb}_{ijklmn} = \frac{1}{2} h_{klpq} {\Rb}_{ij}{}^{pq}{}_{mn}.
\ee
One can also define the Riemann curvature 
$R_{ij}{}^k{}_l(\lam)$ for the induced connection $\G^i{}_j(\lam)$ as usual.

There are two ways to contract a pair of indices on
the area curvature tensor as generalizations of the Ricci tensor:
\be
{\Rb}_{im}{}^{jm}{}_{kl},
\qquad
{\Rb}_{ij}{}^{km}{}_{lm}.
\ee

In terms of the decomposition (\ref{decompose-G}),
we have
\bea
\Rb_{kl}{}^{ij}{}_{mn}
&=& R_{kl}{}^i{}_m(\lam) \d^j_n - R_{kl}{}^j{}_m(\lam) \d^i_n 
- R_{kl}{}^i{}_n(\lam) \d^j_m + R_{kl}{}^j{}_n(\lam) \d^i_m
\nn \\
&&
+ 2D_{[k} \Sb_{l]}{}^{ij}{}_{mn}(\lam)
- \Sb_{k}{}^{ij}{}_{pq}(\lam) \Sb_{l}{}^{pq}{}_{mn}(\lam) 
+ \Sb_{l}{}^{ij}{}_{pq}(\lam) \Sb_{k}{}^{pq}{}_{mn}(\lam).
\eea

For the special case when the area metric is by some metric $g$ (\ref{h=gg}),
the area curvature is determined by $g$ through 
the generalized Levi-Civita connection
with the induced connection given by the Levi-Civita connection of $g$ and $\Sb' = \Kb = 0$.
Then we have
\be
{\Rb}^{ij}{}_{kl} = [R^i{}_k\d^j_l - (i\leftrightarrow j)] - [k\leftrightarrow l],
\ee
where $R^i{}_j$ is the ordinary Riemann tensor (2-form) defined by $g$.
We have
\be
{\Rb}_{ik}{}^{kl}{}_{jl} = (D-2) R_{ij}, 
\label{Ricci-R}
\ee
where $R_{ij}$ is the ordinary Ricci curvature for the metric $g_{ij}$.
(For $D=2$, 
one can always choose $h_{1212}$ to equal to $1$
by general coordinate transformations,
so all 2D spaces are equivalent to the flat 2D space
from the viewpoint of area geometry.)

\subsection{Bianchi identity and generalized Einstein equation}

The Bianchi identity for the area curvature is
\be
D {\Rb} \equiv
d{\Rb} + \Gb \wedge {\Rb} - {\Rb} \wedge\Gb
= 0.
\label{Bianchi-1}
\ee
In the component form, 
it is
\be
D\Rb^{(2)}{}^{pq}{}_{mn} =
\frac{1}{6} dx^i\wedge dx^j\wedge dx^k \; D_{[i} {\Rb}_{jk]}{}^{pq}{}_{mn} = 0,
\label{Bianchi-2}
\ee
where
\be
D_{[i} {\Rb}_{jk]}{}^{pq}{}_{mn} \equiv
\del_{[i} {\Rb}_{jk]}{}^{pq}{}_{mn} + \Gb_{[i}{}^{pq}{}_{rs} {\Rb}_{jk]}{}^{rs}{}_{mn}
- {\Rb}_{[jk}{}^{pq}{}_{rs}\Gb_{i]}{}^{rs}{}_{mn}.
\ee
(The indices $r, s$ in the last two terms are not anti-symmetrized 
together with the indices $i, j, k$.)
As we have explained in Sec.\ref{transf-conn},
when the covariant derivative acts on a differential form,
it ignores the indices to be contracted with $dx$
as a differential form.

In Einstein's theory of gravity,
the Bianchi identity gives us a hint on how to define Einstein's equation 
such that the conservation of energy-momentum tensor is guaranteed.
The situation is different for the gravity theory of the area metric.
The conservation of energy-momentum is a result of 
the invariance of the theory under general coordinate transformations.
In the theory of area metrics,
the gauge symmetry is still merely general coordinate transformations.
We do not expect more conserved quantities than before,
although we do expect more equations of motion 
for more independent components in the area metric.
Therefore, we should not try to define the generalized Einstein equation
from the generalized Bianchi identity as we did in Einstein's theory.

On the other hand,
we wish to find an equation of the area metric
that would reduce to the Einstein equation 
when $h$ is given by a metric $g$ as in (\ref{h=gg}).
Assuming that the area connection is given by $\hat{\Gb}$ 
with both $\Sb$ and $\Kb$ vanishing,
while the induced connection is identical to the Levi-Civita connection
defined by the metric $g$,
we have
\be
\Rb_{kl}{}^{ij}{}_{mn}
= R_{kl}{}^i{}_m \d^j_n - R_{kl}{}^j{}_m \d^i_n 
- R_{kl}{}^i{}_n \d^j_m + R_{kl}{}^j{}_n \d^i_m,
\label{RR}
\ee
where $R_{ij}{}^k{}_l$ is the Riemann tensor for $g$.
Then one can check that the equation 
\be
{\bf G}_{j}{}^{i}{}_{kl} \equiv
{\Rb}_{jm}{}^{im}{}_{kl} 
- \frac{1}{(D-2)} {\Rb}_{jl}{}^{im}{}_{km}
+ \frac{1}{(D-2)} {\Rb}_{jk}{}^{im}{}_{lm}
= 0
\label{GEE}
\ee
reduces to the Einstein equation in vacuum
through the relation (\ref{RR}).
Up to an overall factor,
this is the only combination of the area curvature $\Rb$ 
with this property.
When the source term is present,
the generalized Einstein equation should be of the form
\be
{\Rb}_{jm}{}^{im}{}_{kl} 
- \frac{1}{(D-2)} {\Rb}_{jl}{}^{im}{}_{km}
+ \frac{1}{(D-2)} {\Rb}_{jk}{}^{im}{}_{lm}
= {\bf Z}_{j}{}^{i}{}_{kl},
\label{wave-eq}
\ee
although for the time being we do not know anything about the source term ${\bf Z}$,
except that its trace reduces to something given by the energy momentum tensor $T_{ij}$ as
\be
{\bf Z}_{i}{}^{k}{}_{jk} = (D-1)\left[
T_{ij} - \frac{1}{D-2} g_{ij} T^k{}_k
\right]
\ee
when the area metric is defined by a metric $g$.

\section{Examples of area metric and area connection}
\label{example}

In this section we consider the class of manifolds
with area metrics that can be diagonalized:
\be
h_{AB} = \lam_A \d_{AB},
\label{diag-h}
\ee
where $\lam_A$ is a function of space-time.
We will denote the inverse of $\lam_A$ by
\be
\lam^A \equiv \lam_A^{-1}.
\ee
In this section,
the Einstein summation convention is not used.

There is a universal expression for a candidate of area connection 
which satisfies both the area metricity condition 
and the area torsion-free condition:
\bea
\Gb_{m[ij][kl]} &=&
\frac{1}{2} \del_m \lam_{[ij]} I^{ij}_{kl} 
- \frac{1}{4} \del_k \lam_{[ij]} I^{ij}_{lm} + \frac{1}{4} \del_l \lam_{[ij]} I^{ij}_{km} 
- \frac{1}{4} \del_j \lam_{[im]}I^{im}_{kl} + \frac{1}{4} \del_i \lam_{[jm]} I^{jm}_{kl},
\eea
where 
$I^{ij}_{kl} \equiv \d^i_k \d^j_l - \d^i_l \d^j_k$.
However, 
this expression is not covariant,
and so it is not a good choice for a reference connection.
Hence we shall not use this expression, 
but instead we will find the generalized Levi-Civita connection
by imposing the covariant constraints
(\ref{Gb-expansion-2}) -- (\ref{Gbhat-2}).

Following the procedure outlined in Sec. \ref{GLCC}
to find the area Levi-Civita connection,
we find that the traceless condition of $\Sb'$ (\ref{Sb-1}) gives
\bea
\sum_{k\neq i}\lam^{[ik]}d\lam_{[ik]} - 2(D-2)\G'^i{}_i - 2\sum_{k} \G'^k{}_k = 0,
\\
(D-2)\G'^i{}_j + \Lam^i{}_j \G'^j{}_i = 0 
\qquad \mbox{for} \quad i \neq j,
\eea
where
\be
\Lam^i{}_j \equiv \sum_{k\neq i, j} \lam^{[ik]}\lam_{[jk]}.
\ee
These equations for $\G'^i{}_j$ are solved by
\be
\G'^i{}_j = \frac{1}{2(D-2)}\d^i_j \left[
\sum_{k}\lam^{[ik]}d\lam_{[ik]}
- \frac{1}{2(D-1)}\sum_{kl}\lam^{[kl]}d\lam_{[kl]}
\right].
\ee
Then the area metricity condition implies that $\Sb'$ is 
\be
\Sb'_{ijkl} = \d_{[ij][kl]}\left[
\frac{1}{2} d\lam_{[ij]} - \lam_{[ij]} (\G'^i{}_i + \G'^j{}_j)
\right],
\ee
where $\d_{[ij][kl]}$ is the Kronecker delta $\d_{AB}$ 
with $A = [ij], B = [kl]$.
It follow that
\be
\Gb_{AB} = \frac{1}{2}\d_{AB} d\lam_A + \Kb_{AB}.
\ee

Now we can solve for $\Kb$ by imposing 
the area torsion-free condition.
The Levi Civita area connection is finally given by
\bea
\Gb_{i[ij][ij]} &=& \frac{1}{2} \del_i \lam_{[ij]},
\\
\Gb_{k[ij][ij]} &=& \frac{1}{2} \del_k \lam_{[ij]},
\\
\Gb_{i[ij][jk]} &=& - \frac{1}{4} \del_k \lam_{[ij]},
\\
\Gb_{i[jk][ij]} &=& \frac{1}{4} \del_k \lam_{[ij]},
\eea
where $i, j, k$ are assumed to be all different.
Other components of $\Gb$ vanish.

In the above we have assumed that the area metric is generic.
However,
the solution above for $\G'^i{}_j$ is not unique if
the area metric happens to satisfy the condition
\be
\Lam^i_j \Lam^j_i = (D-2)^2
\label{LamLam}
\ee
for some indices $i \neq j$.
In the subsection below,
we see in an explicit example that this degeneracy in area connection 
somehow does not lead to degeneracy 
in the solutions of the generalized Einstein equation.

\subsection{Static solutions with spherical symmetry}
\label{spherical}

Here we find solutions to the generalized Einstein equation 
for 4-dimensional static diagonalizable area metrics with spherical symmetry.
There is degeneracy (i.e. (\ref{LamLam}) holds) in this case.

Let the space-time coordinates be denoted $(t, r, \theta, \phi)$,
where $t$ will be interpreted as the time coordinate,
$r$ the radial coordinate,
and $\th, \phi$ the angular coordinates.
The spherical symmetry acts on the coordinates $\th, \phi$ 
as it does on the angular coordinates on a 2-sphere.
We assume that there is a coordinate system in which 
all components of the area metric are independent of $t$ (static).
The ansatz of the diagonalizable area metric is
\bea
&h_{trtr} = F_1(r),
\qquad
h_{t\th t\th} = F_2(r),
\\
&h_{t\phi t\phi} = F_2(r)\sin^2\th,
\qquad
h_{r\th r\th} = F_3(r),
\\
&h_{r\phi r\phi} = F_3(r)\sin^2\th,
\qquad
h_{\th\phi\th\phi} = F_4(r)\sin^2\th.
\eea
By a change of coordinates $r \rightarrow r'=r'(r)$,
one can always set 
\be
F_1(r) = -1
\ee
without loss of generality.
We have 4 functions $F_2(r), F_3(r), F_4(r)$ to begin with.
We parametrize them in terms of $\Phi_2(r), \Phi_3(r)$ and $Y(r)$ as
\be
F_2(r) = - e^{\Phi_2(r)},
\qquad
F_3(r) = e^{\Phi_3(r)},
\qquad
F_4(r) = F_2(r)F_3(r)Y(r).
\ee
If $F_4(r) = - F_2(r)F_3(r)$ ($Y = -1$),
this area metric can be interpreted as one that is defined 
by an ordinary metric.

The degeneracy condition (\ref{LamLam}) is satisfied for 
the pairs of indices $(1, 2)$ and $(3, 4)$.
Thus 
$\Gamma'_m{}^1{}_2, \Gamma'_m{}^2{}_1, 
\Gamma'_m{}^3{}_4$ and $\Gamma'_m{}^4{}_3$ satisfy
\bea
\Gamma'_m{}^2{}_1 = -\frac{1}{2}\Lam^2{}_1\Gamma'_m{}^1{}_2,
\\
\Gamma'_m{}^4{}_3 = -\frac{1}{2}\Lam^4{}_3\Gamma'_m{}^3{}_4.
\eea
Demanding spherical symmetry on $\G'_j{}^i{}_k$,
we take the ansatz
\bea
\Gamma'_1{}^1{}_2 = f(r),
\qquad
\Gamma'_4{}^3{}_4 = -\sin\th \cos\th,
\eea
where a new functional degree of freedom $f(r)$ is introduced.
It turns out that the generalized Einstein equation fixes $f(r)$ uniquely,
without introducing more than one solutions for the same area metric.

In addition,
the area metricity condition implies that
\bea
\Gamma'_m{}^i{}_i = \frac{1}{4} \sum_{k\neq i} \lam^{[ik]}\del_m\lam_{[ik]}
- \frac{1}{24} \sum_{k\neq l} \lam^{[kl]}\del_m\lam_{[kl]}
\eea
(without summing over $i$).
All other components of the induced connection
$\Gamma'_j{}^i{}_k$ vanish.

After solving $\Sb'$ and $\Kb$,
we find the resulting area connection $\Gb_{mAB}$ 
($m=1, 2, 3, 4$ and $A, B = 1, 2, \cdots, 6$):
\bea
\Gb_{124} &=& -\Gb_{142} = -\frac{1}{4} e^{\Phi_2} (2 f(r) + \Phi'_2(r)), \\
\Gb_{135} &=& -\Gb_{153} = -\frac{1}{4} e^{\Phi_2} (2 f(r) + \Phi'_2(r)) \sin^2\th, \\
\Gb_{222} &=& - \frac{1}{2} e^{\Phi_2} \; \Phi'_2(r), \\
\Gb_{233} &=& - \frac{1}{2} e^{\Phi_2} \; \Phi'_2(r) \sin^2\th , \\
\Gb_{244} &=& \frac{1}{2} e^{\Phi_3} \; \Phi'_3(r), \\
\Gb_{255} &=& \frac{1}{2} e^{\Phi_3} \; \Phi'_3(r) \sin^2\th, \\
\Gb_{266} &=& - \frac{1}{2} e^{\Phi_2 + \Phi_3}
(Y^\prime(r)+Y(r) (\Phi_2^\prime(r)+\Phi_3^\prime(r)) \sin^2\th, \\ 
\Gb_{312} &=& -\Gb_{321} = - \frac{1}{4} e^{\Phi_2} (2 f(r) - \Phi_2^\prime(r)), \\
\Gb_{333} &=& -\Gb_{423} = \Gb_{432} = - e^{\Phi_2} \cos\th \sin\th, \\
\Gb_{355} &=& -\Gb_{445} = \Gb_{454} = e^{\Phi_3} \cos\th \sin\th, \\
\Gb_{356} &=& -\Gb_{365} = -\Gb_{446} = \Gb_{464} = 
\frac{1}{4} e^{\Phi_2 + \Phi_3}
(Y^\prime(r)+Y(r) (\Phi_2^\prime(r) + \Phi_3^\prime(r)) \sin^2\th, \\
\Gb_{366} &=& - e^{\Phi_2 + \Phi_3} \cos\th \sin\th \; Y(r), \\
\Gb_{413} &=& -\Gb_{431} = - \frac{1}{4} e^{\Phi_2} (2 f(r) - \Phi_2^\prime(r)) \sin^2\th.
\eea
One can check that this area connection is of the form (\ref{Gb-expansion-2})
and satisfies both the area metricity condition and the area torsion-free condition.

The generalized Einstein tensor then gives
the following seven non-trivial components
\bea
{\bf G}_{1}{}^{2}{}_{12} &\equiv& 
\frac{1}{8} e^{ \Phi_2 - \Phi_3 } 
\left[ -4 f^2 + 4 f' + 2 {\Phi'_2}^2 + 2 f \left( \Phi'_2 - \Phi'_3 \right)
- \Phi'_2 \Phi'_3 + 2 \Phi''_2 \right], 
\\
{\bf G}_{1}{}^{3}{}_{13} &\equiv& 
\frac{1}{32 Y} e^{ \Phi_2 - \Phi_3 } \left[
- 4 f^2 Y + 3 Y'_2 \Phi'_2 + 2 f \left( 3 Y' + Y \left( 7 \Phi'_2 - \Phi'_3 \right) \right) 
\right.
\nn \\
&&
\left.
+ Y \left( 16 f' + 8 {\Phi'_2}^2 - \Phi'_2 \Phi'_3 + 8 \Phi''_2 \right)
\right],
\nn
\\
{\bf G}_{2}{}^{1}{}_{12} &\equiv& 
\frac{1}{8} \left[ -4 f' + 3 {\Phi'_2}^2 - \Phi'_2 \Phi'_3
- 2 f \left( \Phi'_2 + \Phi'_3 \right) + 6 \Phi''_2 \right],
\\
{\bf G}_{2}{}^{3}{}_{23} &\equiv& 
-\frac{1}{16 Y^2} \left[
-3 {Y'}^2 + 3 Y \left( Y' \left( 2 \Phi'_2 + \Phi'_3 \right) + 2 Y''_2 \right)
\right.
\nn \\
&&
\left.
+ Y^2 \left( 4 f' + 6 {\Phi'_2}^2 + 2 f \left( \Phi'_2 - 2 \Phi'_3 \right)
+ \Phi'_2 \Phi'_3 + 12 \Phi''_2 + 6 \Phi''_3 \right)
\right],
\\
{\bf G}_{3}{}^{1}{}_{13} &\equiv& 
\frac{1}{32} e^{ \Phi_2 } \left[
- 32 e^{ -\Phi_2 }  + 4 f^2 - 16 f' - Y' \Phi'_2
+ 7 {\Phi'_2}^2 - Y {\Phi'_2}^2 - Y \Phi'_2 \Phi'_3
\right.
\nn \\
&&
\left.
- 2 f \left( Y' + ( 8 + Y ) \Phi'_2 + Y \Phi'_3 \right)
+ 8 \Phi''_2 \right],
\\
{\bf G}_{3}{}^{2}{}_{23} &\equiv& 
\frac{1}{16} e^{ \Phi_2 } \left[
-16 e^{ -\Phi_2 } - 4 f^2 - 4 f'(r) - 2 f \Phi'_2
- 4 Y' \Phi'_2 + 2 {\Phi'_2}^2 - 2 Y {\Phi'_2}^2 - 3 Y' \Phi'_3
\right.
\nn \\
&&
\left. 
- 3 Y \Phi'_2 \Phi'_3 - Y {\Phi'_3}^2 - 2 Y'' + 2 \Phi''_2 - 2 Y \Phi''_2
- 2 Y \Phi''_3 \right],
\\
{\bf G}_{3}{}^{4}{}_{34} &\equiv& 
\frac{1}{32 Y} e^{ \Phi_2 } \left[
-3 {Y'}^2 + Y \left( 32 e^{ -\Phi_2 }  - 4 f^2 - {\Phi'_2}^2 
+ 4 f \left( Y' + \Phi'_2 \right) 
+ 6 Y' \left( 2 \Phi'_2 + \Phi'_3 \right) + 8 Y'' \right)
\right.
\nn \\
&&
\left.
+ Y^2 \left( 7 {\Phi'_2}^2 + 8 \Phi'_2 \Phi'_3 
+ {\Phi'_3}^2 + 4 f \left( \Phi'_2 + \Phi'_3 \right) + 8 \Phi''_2 + 8 \Phi''_3 \right)\right],
\eea
where we omitted redundant relations, e.g.
${\bf G}_{i}{}^{j}{}_{kl} = -{\bf G}_{i}{}^{j}{}_{lk}$, and
${\bf G}_{1}{}^{4}{}_{14}, {\bf G}_{2}{}^{4}{}_{24}, {\bf G}_{4}{}^{1}{}_{14}, {\bf G}_{4}{}^{2}{}_{24}, {\bf G}_{4}{}^{3}{}_{34}$ 
give the same relation as 
${\bf G}_{1}{}^{3}{}_{13}, {\bf G}_{2}{}^{3}{}_{23}, {\bf G}_{3}{}^{1}{}_{13}, {\bf G}_{3}{}^{2}{}_{23}, {\bf G}_{3}{}^{4}{}_{43}$ 
up to an overall factor, respectively.

The generalized Einstein equation in vacuum $\Gb = 0$ gives 7 differential equations 
(not all independent) for the 4 functions $\Phi_2, \Phi_3, Y$ and $f$.
We find two classes of solutions to these equations.

The first class of solutions of
the generalized Einstein equations is given by
\bea
Y(r) &=& -1, \\
f(r) &=& - \frac{1}{2} \Phi'_3(r), \\
\Phi_2(r) &=& \log\left[ (r + c_2)^2 - c_1 \right], \\
\Phi_3(r) &=& \log\left[ (r + c_2)^2 - c_1 \right]
+ c_3 \tanh^{-1}\left[ \frac{r + c_2}{\sqrt{c_1}} \right] + c_4,
\eea
where we need either $c_1 = 0$ or $c_3 = \pm 4$,
while $c_i$'s are all constants.
Note that
all solutions of the area metric with $Y(r) = -1$ can be interpreted as 
the area metric defined by a regular metric with
\be
g_{rr} = - g_{tt}^{-1} = \sqrt{\frac{F_3}{-F_2}},
\qquad
g_{\th\th} = \sqrt{-F_2 F_3},
\qquad
g_{\phi\phi} = \sqrt{-F_2 F_3} \sin^2\th.
\ee
It reduces to the Schwarzschild solution for a mass $m$
when 
\be
c_1 = m^2,
\qquad
c_2 = - m, 
\qquad
c_3 = 4,
\qquad 
c_4 = 0.
\ee

Another class of solutions
allow $Y(r)$ to be an arbitrary negative function.
For an arbitrary negative function $Y(r)$, 
we have
\bea
f(r) &=& - \frac{1}{r-a}, \\
\Phi_2(r) &=& 2 \log (r-a), \\
\Phi_3(r) &=& - \log |Y(r)| - 2 \int^r dr' \; \frac{1}{(r'-a)}\left(1\pm \frac{2}{\sqrt{-Y(r')}}\right).
\eea
As long as $Y \neq -1$,
the area metric can not be interpreted as that defined by an ordinary metric.

Apparently,
there are infinitely many spherically symmetric static solutions of the area metric in vacuum,
in contrast with the no-hair theorem for the Riemannian metric.
We can interpret our result as follows.
In the collapse of a spherical object,
stringy effects become important when 
the energy density is high,
so that the space-time geometry is better described by the area metric.
The abundance of area-metric configurations
may allow the information of the collapsing matter to be preserved
without breaking the spherical symmetry.
It is tempting to speculate a resolution of the information loss paradox
following this line of thoughts.

\section{Summary and outlook}

In this paper we have considered the generalization 
of metric to area metric,
and studied the notion of area connection
as well as area torsion and area curvature.
We propose to explore the possibility that 
the area metric is more appropriate than metric
to describe the geometry of the early universe
in a stage when stringy effects are important.
A phenomenological question is then,
if the metric is inappropriate for describing 
the geometry of the universe at an early stage,
why is it suitable to describe our present-day universe?
We have shown in Sec. \ref{effective-metric}
that there are potential energy terms which 
can drive the area metric towards a configuration 
that admits an approximation through (\ref{h=gg}) for a certain metric.

We have pointed out the fact that 
there are many covariant tensorial degrees of freedom 
in the area connection 
and there can be many ways to fix them.
We found a class of 3-parameter area connections 
to satisfy the area metricity condition 
and the area torsion-free condition,
and we pointed out that the algebraically simplest choice 
is $(\lam = 0, \alpha = 0, \beta =1)$.
In this case,
when the area metric is given by a metric $g$ through (\ref{h=gg}),
the area connection can be taken as $\Gb = \hat{\Gb}'$,
with $\Gamma'$ given by the Christoffel symbol of the metric $g$.
The generalized Einstein equation is defined so that 
it reduces to the Einstein equation in this situation.

One may hope to impose more constraints
on the area connection,
such as the covariant constancy of the volume form.
The volume form is automatically covariantly constant 
in Riemannian geometry if the metricity condition is satisfied.
But the area metricity condition is insufficient,
contrary to what was claimed in some of the literatures.
In fact, 
the covariant constancy of the volume form 
with respect to the area connection
\footnote{
Recall that we have shown in Sec. \ref{InducedConnection}
that the volume form is always covariantly constant
with respect to the induced connection $\G^i{}_j(0)$.
}
imposes a very strong constraint on the area connection,
which in general constrains the area metric (and its derivatives).
We will explain this in Appendix \ref{app}.

A problem with our formulation of the gravity of area metric 
presented above is that 
it does not seem to admit an action principle.
With an action principle,
the current that couples to the area metric $h_{ijkl}$
is expected to be a tensor ${\bf Z}^{ijkl}$
with 4 upper indices
(or 4 lower indices by using the area metric), 
unlike the tensor ${\bf Z}_i{}^{jkl}$ in (\ref{wave-eq}).
Similarly,
the tensor $\Rb_{ij}{}^{kl}{}_{mn}$ does not admit 
the definition of scalar curvature through the contraction of indices,
if the only additional tensor available is the area metric.
A possibility is that the action principle for the area metric theory
is available only in certain dimensions when
the volume form can be used to do the trick.
For example, 
in 6 dimensions,
one can define ${\bf Z}_{ijkl} \equiv {\bf Z}_{i}{}^{mnp}\Omega_{mnpjkl}$.
We leave the issue of action principle formulation 
of the area gravity for the future.

A related issue that has been left out above 
is the description of matters in the background of
a space-time geometry defined by an area metric.
As we mentioned in the Introduction, 
the Yang-Mills action 
(including the Maxwell action)
is naturally defined by the area metric.
The propagation of light in an area metric background
has been studied in the literature \cite{Punzi:2007di},
and the causal structure for a given metric 
is analyzed in detail \cite{Punzi:2007di,Schuller:2009hn}
for 4 dimensional space-time.
The quantization of general linear electrodynamics
has also been considered \cite{Rivera:2011rx}.

It is less clear how to describe the motion of point masses 
in a background defined by the area metric.
It turns out that the area metric 
determines an effective Finsler geometry 
and a point particle moves along its geodesics
\cite{Punzi:2009yq}.
It will also be interesting to consider
higher dimensional branes in the geometry 
defined by area metric.

Generalizations of area metric to metrics of higher dimensional volumes 
can be studied in a similar fashion.
The metric for a $d$-dimensional volume can be defined as
\be
dv^2 = h_{i_1 \cdots i_d j_1 \cdots j_d} 
(dx^{i_1}\wedge\cdots\wedge dx^{i_d})\otimes
(dx^{j_1}\wedge\cdots\wedge dx^{j_d}),
\ee
where the volume metric $h$ should be 
totally anti-symmetrized in $(i_1, \cdots, i_d)$,
and totally anti-symmetrized in $(j_1 \cdots, j_d)$.
It should also satisfy the cyclicity condition
\be
\sum_{cylic(j)}
h_{i_1\cdots i_{d-1}[j_1\cdots j_{d+1}]} = 0,
\ee
which is summed over cyclic permutations of $(j_1, \cdots, j_{d+1})$.

Finally, 
a generalization of the area metric analogous to 
the generalization of the usual metric to Finsler geometry \cite{Finsler} is possible.
Let us first review the notion of Finsler geometry.
For the action of a particle 
\be
S = \int d\tau \; L(x, \dot{x})
\ee
with reparametrization symmetry,
the Lagrangian must be homogeneous of degree $1$
in $\dot{x} \equiv dx/d\tau$:
\be
L(x, \lam \dot{x}) = \lam L(x, \dot{x}).
\ee
Then it makes sense to define,
up to an overall factor, 
the length of an infinitesimal line element as
\be
ds = L(x, dx).
\ee
Finsler geometry is the geometry equipped 
with this class of definitions of length.
Similarly,
we can define the area Finsler geometry by
a generalized string action
\be
S = \int d\tau d\sigma \; {\cal L}(x, \dot{x}, x'),
\ee
where the Lagrangian density should satisfy 
\bea
{\cal L}(x, \lam\dot{x}, x') = {\cal L}(x, \dot{x}, \lam x')
= \lam{\cal L}(x, \dot{x}, x').
\eea
It is then consistent to define the notion of area via
\be
da = 
d\tau d\sigma \; {\cal L}(x, \dot{x}, x')
= {\cal L}(x, d\tau \dot{x}, d\sigma x').
\ee
Such a generalized notion of area,
and more generally the volume of $m$-dimensional submanifolds 
embedded in an $n$-dimensional space defined analogously,
have been considered under the terminology of ``areal geometry'' \cite{Areal}.
In these considerations,
the metric and connection in general depend not only on $x$
but also on the derivatives of $x$ with respect to world-volume coordinates.
It will be interesting if any of these generalized notion of geometry will 
find its natural applications in string theory through extended objects like D-branes.
We leave this possibility for future study.
For a recent work in this direction, 
see \cite{Ootsuka:2014hpa}.

\section*{Acknowledgement}

The authors would like to thank 
Heng-Yu Chen, Chong-Sun Chu, Kazuo Hosomichi,
Hikaru Kawai, Yutaka Matsuo and Shu-Heng Shao
for their interest and discussions.
The work is supported in part by
Ministry of Science and Technology, Taiwan, R.O.C.
and by National Taiwan University.

\appendix

\section{Covariant constancy of the volume form}
\label{app}

In Riemannian geometry,
the volume form is covariantly constant
as a result of the metricity condition.
However,
the volume form (\ref{Omega}) defined from an area metric
is in general not covariantly constant with respect to the area connection,
even when the area metricity condition is satisfied.

If one wants to impose the covariant constancy of the volume form
\be
(D\Omega)_{i_1\cdots i_D} = 0
\label{DOmega}
\ee
as a constraint on the area connection,
an immediate problem is that
the covariant derivative with respect to 
the area connection $\Gb^{ij}{}_{kl}$ 
cannot be straightforwardly defined in odd dimensions.

For $D = $ even, 
one can interpret the covariant constancy of 
the volume form (\ref{DOmega}) as
\be
d\Omega_{i_1\cdots i_D} 
- \frac{1}{2} \Omega_{j_1 j_2 i_3\cdots i_D}
\Gb^{j_1 j_2}{}_{i_1 i_2}
- \cdots
- \frac{1}{2} \Omega_{i_1\cdots i_{D-2} j_{D-1} j_D}
\Gb^{j_{D-1} j_D}{}_{i_{D-1} i_D} = 0.
\label{DO=0}
\ee
This condition (\ref{DO=0}) is highly non-trivial,
and it is too restrictive to be imposed on the area connection. 
The origin of the problem is that,
although the first term in (\ref{DO=0}) is
totally anti-symmetrized in all indices,
the rest of the terms are {\em a priori} not.

As an example,
in 4 dimensions,
the condition (\ref{DO=0}) implies 
all of the following equations
(without summing over repeated indices)
\bea
d\omega\eps_{ijkl}&=&
\omega\eps_{ijkl}(\Gb^{ij}{}_{ij}+\Gb^{kl}{}_{kl}),
\\
d\omega\eps_{ijik}&=&0=
\omega(\eps_{jlik}\Gb^{jl}{}_{ij}+\eps_{ijkl}\Gb^{kl}{}_{ik}),
\\
d\omega\eps_{ijij}&=&0=
2\omega\eps_{klij}\Gb^{kl}{}_{ij},
\eea
for an arbitrary permutation $\{i,j,k,l\}$ of $\{0,1,2,3\}$.
They constitute a total of $84$ constraints on the area connection.
(They are not all linearly dependent -- 
see (\ref{deth=G}) below.)

The number of constraints derived from (\ref{DO=0}) increases
with the dimension $D$ much faster than 
that of components of the area connection.
Therefore,
in general the condition of covariant constancy of the volume form (\ref{DO=0}) imposes
too many constraints to be solved together with the area metricity condition.
It is therefore unlikely to impose (\ref{DO=0}) for all $D$,
unless further constraints are imposed on the area metric
so that there is more linear dependence in the constraints (\ref{DO=0}).

The fact that the volume form is in general not covariantly constant 
means that the operation of Hodge dual 
is in general not commutative with the operation of covariant derivatives,
with respect to the area connection.
(On the other hand, 
as we have seen in Sec. \ref{InducedConnection},
the volume form is covariantly constant with respect to the induced connection.)

There is on the other hand a tensor whose covariant constancy is guaranteed
by the area metricity condition.
It is
\be
d(\det h) \eps_{A_1 A_2 \cdots A_N} = 
(\det h) \eps_{B A_2 \cdots A_N}\Gb^B{}_{A_1}
+ (\det h) \eps_{A_1 B \cdots A_N}\Gb^B{}_{A_2}
+ (\det h) \eps_{A_1 A_2 \cdots B}\Gb^B{}_{A_N},
\ee
where $A_i, B = 1, 2, \cdots, N\equiv C^D_2$,
and the determinant $(\det h)$ is defined in (\ref{deth}).
This equation is equivalent to 
\be
d\log(\det h) = \Gb^A{}_A,
\label{deth=G}
\ee
which can be derived as a result of the area metricity condition.

\end{CJK} 
\end{document}